# Giant 2D Skyrmion Topological Hall Effect with Ultrawide Temperature Window and Low-Current Manipulation in 2D Room-Temperature Ferromagnetic Crystals

*Gaojie Zhang*[1,2], *Qingyuan Luo*[3], *Xiaokun Wen*[1,2], *Hao Wu*[1,2,*], *Li Yang*[1,2], *Wen Jin*[1,2], *Luji Li*[1,2], *Jia Zhang*[4], *Wenfeng Zhang*[1,2,5], *Haibo Shu*[3,*], *Haixin Chang*[1,2,5,*]

[1]State Key Laboratory of Material Processing and Die & Mold Technology, School of Materials Science and Engineering, Huazhong University of Science and Technology, Wuhan 430074, China.

[2]Wuhan National High Magnetic Field Center and Institute for Quantum Science and Engineering, Huazhong University of Science and Technology, Wuhan 430074, China.

[3]College of Optical and Electronic Technology, China Jiliang University, Hangzhou 310018, China.

[4]School of Physics and Wuhan National High Magnetic Field Center, Huazhong University of Science and Technology, 430074 Wuhan, China.

[5]Shenzhen R&D Center of Huazhong University of Science and Technology, Shenzhen 518000, China.

[*]Corresponding authors. E-mail: hxchang@hust.edu.cn, h_wu@hust.edu.cn, shuhaibo@cjlu.edu.cn

Keywords: giant topological Hall effect, ultrawide temperature window, magnetic skyrmions, two-dimensional ferromagnetic crystals, room temperature

**The discovery and manipulation of topological Hall effect (THE), an abnormal magnetoelectric response mostly related to the Dzyaloshinskii-Moriya interaction (DMI), are promising for next-generation spintronic devices based on topological spin textures such as magnetic skyrmions. However, most skyrmions and THE are stabilized in a narrow temperature window either below or over room temperature with high critical current manipulation. It is still elusive and challenging to achieve large THE with both wide temperature window till room temperature and low critical current manipulation. Here, by using controllable, naturally-oxidized, sub-20 and sub-10 nm 2D van der Waals room-temperature ferromagnetic $Fe_3GaTe_{2-x}$ crystals, robust 2D THE with ultrawide temperature window ranging in three orders of magnitude from 2 to 300 K is reported, combining with giant THE of ~5.4 μΩ·cm at 10 K and ~0.15 μΩ·cm at 300 K which is 1-3 orders of magnitude larger than that of all known room-temperature 2D skyrmion systems. Moreover, room-temperature current-controlled THE is also realized with a low critical current density of ~$6.2\times10^5$ $A\cdot cm^{-2}$. First-principles calculations unveil natural oxidation-induced highly-enhanced 2D interfacial DMI reasonable for robust giant THE. This work paves the way to room-temperature, electrically-controlled 2D THE-based practical spintronic devices.**

# 1. Introduction

Over the past decade, realizing real-space topological spin textures at room temperature has attracted enormous attention for topological physics and spintronics[1-6]. Understanding the abnormal magnetoelectric transport behavior of carriers coupled with these topological spin textures is crucial for the practical application of spintronic devices. In general, topological spin textures with scalar spin chirality generate an effective magnetic field that results in the topological Hall effect (THE) when the carriers passing through it, thereby linking the local topological spin textures to an electrical response[7-9]. The THE appears as an antisymmetric spike in the magnetoelectric transport image, which has been observed in some magnetic systems with inversion symmetry breaking and strong spin-orbit coupling (SOC)[6-13]. Up to now, THE has become an effective tool for electrical detection and manipulation of topological spin textures such as magnetic skyrmions[6-9, 14-16]. However, most skyrmions and THE can only be stabilized in a narrow temperature window either below or over room temperature with high critical current manipulation[6, 9, 14]. It is still elusive and challenging to realize large THE with both wide temperature window till room temperature and low critical current manipulation.

The THE-hosting magnetic systems mostly possess the appreciable Dzyaloshinskii-Moriya interaction (DMI), $H_{DMI}=-D_{ij}\cdot(S_i\times S_j)$, which implies an antisymmetric exchange interaction between two neighboring spins $S_i$ and $S_j$ with strength and direction determined by Moriya vector $D_{ij}$[9-11]. Compared with bulk DMI induced THE in chiral, kagome, and frustrated magnets[7, 17-19], interfacial DMI induced THE in 2D heterostructures or superlattices provide a broader platform for the miniaturization, integration and controllability of 2D spintronic devices, which is more desirable in current microelectronic industry[8-12]. However, 2D interfacial DMI is mostly induced in conventional heavy-metal/ferromagnet ultrathin heterostructures so far[2, 9, 11], and delicate control of 2D DMI for 2D THE manipulation at room temperature is still challenging.

In this work, by natural oxidization of sub-20 and sub-10 nm 2D van der Waals (vdW) room-temperature ferromagnetic $Fe_3GaTe_{2-x}$ crystal (O-$Fe_3GaTe_{2-x}$), robust 2D THE with an ultrawide temperature window ranging in three orders of magnitude from 2 to 300 K is demonstrated in O-$Fe_3GaTe_{2-x}$/$Fe_3GaTe_{2-x}$ (O-FGaT/FGaT) heterostructures. Remarkably, 2D O-FGaT/FGaT exhibits giant THE with topological Hall resistivity ($\rho_{TH}$) of ~5.4 μΩ·cm at 10 K and ~0.15 μΩ·cm at 300 K which is 1-3 orders of magnitude better than that of all known room-temperature 2D skyrmion systems. Moreover, the current-controlled THE in 2D O-FGaT/FGaT at room temperature reveals a low critical current density of ~$6.2\times10^5$ $A\cdot cm^{-2}$. Using first-principles calculations, we unveil that the surface natural oxidation induces O-Fe, O-Ga, O-Te orbital hybridization and surface charge transfer which further induce ~8.2-26.3 times highly enhanced, sizeable interfacial DMI in 2D O-FGaT/FGaT, which provides an opportunity to engineer 2D vdW heterostructures for robust large THE-based spin memory with low-critical-current tunability.

## 2. Results

### 2.1. Characterizations of pristine and oxidized 2D vdW room-temperature ferromagnetic $Fe_3GaTe_{2-x}$

The pristine vdW-stacked $Fe_3GaTe_{2-x}$ crystal is a layered metallic ferromagnet with $P6_3/mmc$ space group and some Te vacancies. As shown in **Figure 1a**, a $Fe_3Ga$ heterometallic slab is sandwiched between two Te (or Te vacancies) layers, forming a monolayer $Fe_3GaTe_{2-x}$ stacked along the *c*-axis. The crystal structure of $Fe_3GaTe_{2-x}$ is basically similar to that of $Fe_3GaTe_2$[20] and confirmed by X-ray diffraction (XRD), high-resolution transmission electron microscopy (HRTEM), and energy-dispersive spectroscopy (EDS) (**Figure S1 and S2, Supporting Information**). The uniform elemental distribution of Fe, Ga, Te and atomic ratio of 3.12:0.95:1.70 are identified, implying the existence of Te vacancies in the pristine $Fe_3GaTe_{2-x}$ crystal. This speculation is further verified by electron probe micro-analyzer (EPMA) with Te vacancy content of ~15 at% (**Figure S3, Supporting Information**). The naturally oxidized $Fe_3GaTe_{2-x}$ nanosheet is examined by a high-angle annular dark-field scanning transmission election microscopy (HAADF-STEM) along the [120] direction (**Figure 1b**). **Figure 1c-e** clearly show the vdW-stacked structure and the presence of Te vacancies with atomic resolution in pristine $Fe_3GaTe_{2-x}$. The natural oxidation leads to the formation of O-FGaT/FGaT heterostructure with an oxidation layer thickness of ~4.5 nm, which can be examined by EDS mappings (**Figure 1f**). Moreover, the 2D O-FGaT layer is also confirmed by the element line distribution along the depth direction (**Figure 1g**) and $Ar^+$-etched X-ray photoelectron spectroscopy (XPS) (**Figure S4**, more discussions in **Note S1**, **Supporting Information**).

### 2.2. Giant 2D THE with ultrawide temperature window

The pristine bulk $Fe_3GaTe_{2-x}$ crystals and its 2D nanosheet host an above-room-temperature ferromagnetism (Curie temperature $T_C$≈350-360 K) and large perpendicular magnetic anisotropy (PMA) (**Figure S5 and S6**, more discussions in **Note S2, Supporting Information**). Only typical anomalous Hall effect (AHE) exists in pristine $Fe_3GaTe_{2-x}$ 2D nanosheets and no THE is observed (**Figure S6c, Supporting Information**). The anomalous Hall device geometry for O-FGaT/FGaT heterostructure is shown in **Figure 2a**, and the thickness of as-tested 2D O-FGaT/FGaT is ranged from 77 down to 9.8 nm (**Figure S7, Supporting Information**). As shown in **Figure 2b,c**, 77 and 30 nm O-FGaT/FGaT heterostructures exhibit obvious AHE hysteresis loops when the measured temperature is lower than ~370 K. When the thickness of O-FGaT/FGaT is less than 20 nm, the unusual dips and peaks emerge around the coercivity ($H_C$) and disappear at higher fields, indicating that the THE is induced when the magnetic moment start to be reversed (**Figure 2d,e** and **Figure S8c, Supporting Information**). Note that the AHE and THE in the sub-20 nm O-FGaT/FGaT can remain up to room temperature, implying the coexistence of robust ferromagnetic order and topological spin textures in 2D scale at room temperature. Moreover, the normalized anomalous Hall resistivity ($\rho_{AH}$) and $H_C$ show a thickness dependence in these O-FGaT/FGaT with $T_C$ ~330-370 K decreasing with reducing thickness (**Figure 2f**,**g**).

Before discussing the 2D THE in sub-20 nm O-FGaT/FGaT, it is necessary to review some recent criticisms about the artifact "THE"[21]. Based on recently-reported distinguishing methods[22, 23], the possibility of artifact "THE" has been discussed and eliminated (more discussions in **Note S3** and **Figure S9, Supporting Information**), thereby implying that the observed 2D THE may originate from topologically nontrivial spin textures such as magnetic skyrmions[10, 13]. Actually, the observed 2D THE in sub-20 nm O-FGaT/FGaT is most likely major induced by Néel-type skyrmions stabilized by interfacial DMI, similar with other Néel-type skyrmion systems such as $CrTe_2/Bi_2Te_3$[12], $WTe_2/Fe_3GeTe_2$[10], and O-$Fe_3GeTe_2/Fe_3GeTe_2$[24, 25]. This deduction is confirmed by Lorentz-TEM at room temperature, as the skyrmion in 2D O-FGaT/FGaT is only visible at non-zero tilt angle while invisible at zero tilt and reverse its contrast when reversing the tilt angle or focus direction (**Figure S10, Supporting Information**), consistent with the nature of Néel-type skyrmions[10, 24, 25]. A thinner 2D O-FGaT/FGaT shows a higher skyrmion density (**Figure S11, Supporting Information**). In contrast, no skyrmions are observed in the pristine non-oxidized thin 2D $Fe_3GaTe_{2-x}$ nanosheet (**Figure S12, Supporting Information**), implying the important role of surface oxidization for realizing Néel-type skyrmions in this 2D system.

Hence, Hall effect can be decomposed into three terms, including ordinary Hall effect (OHE), AHE and THE (more discussions in **Note S4**, **Supporting Information**). In order to study the temperature and magnetic field dependence of THE, the linear fitting and a step function $M_0 \tanh\left[\frac{B}{a_0}\text{-}H_0\right]$, where $M_0$, $a_0$ and $H_0$ are fitting parameters, are used to single out the OHE and AHE contributions[12] (**Figure 2h** and **Figure S13, Supporting Information**). The skyrmion phase diagrams for 9.8, 13 and 16 nm O-FGaT/FGaT are plotted based on temperature-dependent THE (**Figure 3a-d** and **Figure S14, Supporting Information**). In certain $\rho_{TH}$-B curves, the occurrence of both positive and negative peaks near the $H_C$ is attributed to the nucleation and pinning of skyrmions, similar to other skyrmion systems[26]. Moreover, as temperature increases, the 2D THE exists in a smaller magnetic field and gradually weakens but persists at room temperature. The large THE at low temperatures implies a stronger effective magnetic field ($B_{eff}$) induced by magnetic skyrmions in real space[8, 12]. For 19 nm O-FGaT/FGaT, the emergent THE exists in an ultrawide temperature window ranging in three orders of magnitude from 2 to 300 K (**Figure 3e**), suggesting the robust skyrmions in these sub-20 nm O-FGaT/FGaT. At room temperature, the variation of $\rho_{TH}$ with magnetic field of this 19 nm O-FGaT/FGaT has a similar variation trend to that of skyrmions density with magnetic field of Lorentz-TEM thin sample mentioned above, which further confirms the correlation between THE and skyrmions to some extent (**Figure S15, Supporting Information**). Moreover, although spin chirality fluctuations may also contribute to the $\rho_{TH}$ together with skyrmions in some THE-hosting systems, there is no direct experimental evidence supports this possibility in the 2D O-FGaT/FGaT system so far. Therefore, according to current results and analysis, the most likely major origin of THE in thin 2D O-FGaT/FGaT is the skyrmions.

The magnitude of 2D THE presents some thickness dependence. Remarkably, the 13 nm O-FGaT/FGaT shows giant $\rho_{TH}$ as large as ~5.4 μΩ·cm at 10 K and ~0.15 μΩ·cm at 300 K (**Figure 3a** and **Table S1, Supporting Information**), respectively, reflecting the large coupling strength between current and skyrmions[17]. Note the room-temperature $\rho_{TH}$ in 2D O-FGaT/FGaT is 1-3 orders of magnitude better as compared with all other 2D skyrmion systems such as $[Ir/Fe/Co/Pt]_{20}$ (0.03 μΩ·cm)[11], $Tm_3Fe_5O_{12}/Pt$ (0.0046 μΩ·cm)[9], and $[Co/Pt]_5$ (0.01 μΩ·cm)[27] (**Figure 3f** and **Table S2, Supporting Information**). For 19, 16 and 13 nm O-FGaT/FGaT, $\rho_{TH}$ increases with reduced thickness (**Table S1, Supporting Information**). But $\rho_{TH}$ of 9.8 nm O-FGaT/FGaT deteriorates with worst performance (**Table S1, Supporting Information**), which may originate from the oxidation-induced structural degradation due to the fact that thin sample is more likely to be damaged by air oxidation compared with thicker one. Further, the THE-derived skyrmion size ($n_{sk}^{-1/2}$) of four sub-20 nm O-FGaT/FGaT are summarized in **Table S1** (more discussions in **Note S4, Supporting Information**), and the 13 nm O-FGaT/FGaT shows skyrmion size of 1.2 and 12.3 nm at 10 and 300 K, respectively. Such extremely tiny THE-derived skyrmion size is the smallest one for 2D skyrmion systems[8, 18] (**Figure S16, Supporting Information**). Overall, this robust 2D THE combining an ultrawide temperature window till room temperature and giant $\rho_{TH}$ in 2D O-FGaT/FGaT is superior to that of all known 2D skyrmion systems (**Figure 3g** and **Table S2, Supporting Information**), which is essential for practical applications of 2D THE-based spintronic devices.

### 2.3. Room-temperature low-current manipulation of THE

Up to now, current-controlled THE has been studied theoretically and experimentally in many skyrmion systems, which provides a potential possibility for electrical detection of skyrmion motion[6, 14-16, 28]. Hence, we perform the current-density dependence of 2D THE in 19 and 13 nm O-FGaT/FGaT at room temperature (**Figure 4**). Prior to this, the influence of joule heating effect is evaluated by $\rho_{xx}$ and saturated $\rho_{AH}$ at each current density (**Figure S17**, more discussions in **Note S5, Supporting Information**). With the increase of current density, the $\rho_{TH}$ of 19 and 13 nm O-FGaT/FGaT first remains unchanged and then gradually decreases, as shown in **Figure 4a,b**. In order to confirm whether the decrease of $\rho_{TH}$ is due to the skyrmions motion, we use some formulas to analyze these THE data. In the system of current-driven skyrmion motion, the current-density dependence of the $\rho_{TH}$ follows the formula[14]: $\rho_{TH}(j)=\rho_{TH}(j\leq j_c)[1-A\left(1-{j_c}/{j}\right)]$ , where $\rho_{TH}(j\leq j_c)=PR_0B_{eff}=PR_0n_{sk}\emptyset_0$ which has been mentioned in **Note S4, Supporting Information**, A is an introducing coefficient and $j_c$ is the critical current density. After nonlinear fitting, the current-density dependent $\rho_{TH}$ data are in good agreement with the relation $\rho_{TH}(j>j_c)\propto 1/j$ (upper panels in **Figure 4c,d**), implying a potential possibility of current-driven skyrmion motion in 2D O-FGaT/FGaT. Actually, since skyrmions have been proved to induce exactly one quantum of emergent magnetic flux per skyrmion[29], a moving skyrmion will produce an emergent electric field perpendicular to the direction of the skyrmion motion according to the Faraday's law of induction. This emergent electric field opposes to the topological Hall field arising from the static skyrmions, thereby suppressing the $\rho_{TH}$[14, 28]. Meanwhile, the $j_c$ for current-controlled THE in 19 and 13 nm O-FGaT/FGaT are ~$6.2\times10^5$ A·cm$^{-2}$ and ~$7.82\times10^5$ A·cm$^{-2}$, respectively, 1-2 orders of magnitude lower than that of most room-temperature 2D skyrmion systems[30-32] (**Table S3, Supporting Information**).

Furthermore, the skyrmions drift velocity ($v_d$) is also linearly fitted by using formula[6, 14]: $v_d=jR_0[1-\rho_{TH}(j)/\rho_{TH}(j\leq j_c)]$ and shows good fitting quality (lower panels in **Figure 4c,d**). For 19 and 13 nm O-FGaT/FGaT, the typical $v_d$ are calculated as ~0.54 and ~0.82 m·s$^{-1}$, four orders of magnitude larger than that of some heavy metal/ferromagnet heterostructures such as Ta/CoFeB/$TaO_x$ ($v_d$=$2.5\times10^{-5}$ m·s$^{-1}$)[2]. Remarkably, 2D O-FGaT/FGaT exhibits low $j_c$ and moderate $v_d$ at room temperature compared with other 2D skyrmion systems, which are crucial for practical application of 2D skyrmion-based logic and memory devices (**Figure 4e**). Also, the combination of low current and THE measurements offers fundamental insights into the emergent electrodynamics of skyrmions motion, which will be important for practical applications in the long term[15, 28]. Nevertheless, given the complexity and challenge of fully identifying the underlying physical nature behind the electrically-controlled THE and skyrmions, more studies in this promising 2D room-temperature ferromagnetic crystals are necessary in the future.

## 2.4. First-principles calculations of DMI

The first-principles calculations are used to reveal the surface oxidation effect on DMI in O-FGaT/FGaT (see **Experimental Section**). $Fe_3GaTe_{2-x}$ shows higher activity and affinity to incorporate with oxygen than perfect $Fe_3GaTe_2$ under ambient condition, confirming the possibility to induce the formation of a uniform oxide layer on the surface of $Fe_3GaTe_{2-x}$ by the exposure to air (**Table S4, Supporting Information**). The DMI in bulk $Fe_3GaTe_{2-x}$ and bilayer $Fe_3GaTe_{2-x}$ with/without surface oxidization are obtained by calculating the layer-resolved SOC energy difference ($\Delta E_{SOC}$), microscopic and micromagnetic DMI parameters (*d* and *D*) (**Figure 5 and Figure S18, Supporting Information**). In all cases, non-zero total DMI energies ($|D|\neq 0$) are attributed to asymmetric crystal structure caused by Te vacancies and/or surface oxidization. However, the THE-hosting magnetic systems usually present an appreciable $|D|$ up to ~1-2 $mJ\cdot m^{-2}$, such as $WTe_2/Fe_3GeTe_2$ and $Tm_3Fe_5O_{12}/Pt$[9, 10]. Given that the bulk transport is dominant in thick O-FGaT/FGaT samples, we evaluate the DMI in the bulk $Fe_3GaTe_{2-x}$ and find a weak $|D|$ of ~0.379 $mJ\cdot m^{-2}$ (**Figure 5a**), which is difficult to generate the detectable THE. This result also supports the absence of THE in O-FGaT/FGaT when their thickness over 30 nm. For pristine bilayer $Fe_3GaTe_{2-x}$, a slight crystal asymmetry caused by Te vacancies produces a weak $|D|$ of ~0.28 $mJ\cdot m^{-2}$, resembling to the magnitude of that of bulk $Fe_3GaTe_{2-x}$ (**Figure 5b**). In contrast, an additional large DMI contribution is recognized when introducing the surface O atoms on the bilayer $Fe_3GaTe_{2-x}$ (**Figure 5c,d**). The oxidation-induced large $|D|$ for the O-substituted and O-interstitial bilayer $Fe_3GaTe_{2-x}$ reaches ~7.354 and ~2.301 $mJ\cdot m^{-2}$, which is ~26.3 and ~8.2 times that of pristine case, respectively (**Figure 5e**). The significant enhancement of DMI in bilayer $Fe_3GaTe_{2-x}$ with surface oxidation supports the existence of robust THE in 2D O-FGaT/FGaT heterostructures. Further calculations show that introducing surface O atoms in 2D $Fe_3GaTe_{2-x}$ crystals will cause the shift of total, Fe-3d, Ga-4p, Te-5p states toward the low-energy direction (**Figure S19, Supporting Information**) and the obvious charge transfer from other adjacent atoms into O atoms (**Table S5, Supporting Information**). These results imply that the orbital hybridization between O-2p and other atoms (including Fe-3d, Ga-4p, Te-5p) and surface charge transfer play an essential role in sizeable DMI of 2D O-FGaT/FGaT heterostructures[33-35].

## 3. Conclusion

In summary, with reliable naturally-oxidized interfaces and highly enhanced large 2D DMI, the robust and giant 2D THE with ultrawide temperature window ranging from 2-300 K and low $j_c$ for room-temperature current-controlled THE is reported in down to sub-20 and sub-10 nm 2D vdW room-temperature ferromagnet-based heterostructures. This 2D THE-hosting room-temperature ferromagnetic system has the unique advantage of balance in giant $\rho_{TH}$, ultrawide temperature window till room temperature and low $j_c$, which shows great potential in the real practical room-temperature applications of 2D skyrmion systems. Natural surface oxidization compatible with industrial semiconductor processing may provide general methodology to tune 2D DMI for spin transport control in 2D ferromagnetic crystals. This work not only proves the earth-abundant light element of oxygen can induce giant 2D THE much better than that from heavy metals and strong SOC compounds, but also paves the avenue to electrical control of room-temperature 2D THE and skyrmions for 2D topological and spintronic devices.

## 4. Experimental Section

*Crystal growth, natural oxidization and device fabrication:* $Fe_3GaTe_{2-x}$ single crystals were grown by a self-flux method similar to a previous report[20]. The laser direct writing machine (MicroWriter ML3, DMO) and e-beam evaporation (PD-500S, PDVACUUM) were used to fabricate the Hall bar electrodes (Cr/Au: 8/17 nm) on the $SiO_2$/Si substrate. A mechanically-exfoliated $Fe_3GaTe_{2-x}$ nanosheet was transferred onto the Hall bar electrodes through the polydimethylsiloxane (PDMS)-assisted dry transfer method in an argon-filled glove box ($H_2O$, $O_2$<1 ppm). To realize natural oxidization in exfoliated $Fe_3GaTe_{2-x}$ single nanosheets, the nanosheets were naturally oxidized in the air for 20-30 min, forming the O-FGaT/FGaT heterostructures.

*Crystal characterizations:* The morphology, thickness, structure and elements of $Fe_3GaTe_{2-x}$ crystals were characterized at room temperature by Optical microscopy (OM, MV6100), powder X-ray diffraction (XRD, Smartlab SE, Rigaku Corporation) with Cu Kα radiation (λ=0.154 nm), atomic force microscopy (AFM, Dimension EDGE, Bruker), X-ray photoelectron spectroscopy (XPS, AXIS SUPRA+, Shimadzu), electron probe micro-analyzer (EPMA, 8050G, Shimadzu), field-emission transmission electron microscopy (FTEM, Tecnai G2 F30, FEI), and spherical aberration correction transmission electron spectroscopy (ACTEM, Themis Z, FEI) equipped with energy-dispersive spectroscopy (EDS). Cross-sectional ACTEM specimens were prepared on the h-BN/O-FGaT/FGaT heterostructures, using a focused ion beam instrument (FIB, Helios 5 CX, FEI). To protect the surface from ion beam damage, the platinum layer before milling was deposited. The magnetic properties of $Fe_3GaTe_{2-x}$ crystals were measured by a vibrating sample magnetometer (VSM) module in physical property measurement system (PPMS DynaCool, Quantum Design, USA).

*Magneto-transport measurements:* The magneto-transport measurements were performed in a physical property measurement system (PPMS, DynaCool, Quantum Design). The magnetic field was applied perpendicular to the sample unless otherwise stated. Specially, the magnetic field was changed with the interval of 20-100 Oe around the dips and peaks regime. Each data was tested 25 times in R-T and R-B curves for an average with the constant current mode. The constant current for measurement was set at 1-10 μA depending on the resistance of devices. Additional millisecond current pulses were used to test the current-controlled 2D THE at room temperature. Waiting several seconds after each pulse and then read the resistance.

*Lorentz-TEM measurements:* The as-tested O-FGaT/FGaT and pristine non-oxidized $Fe_3GaTe_{2-x}$ nanosheets were placed on a 50 nm amorphous silicon nitride substrate and capped with the 8 nm Au film to avoid further oxidization. Then, the in situ skyrmions and magnetic domains imaging at room temperature were carried out by using a 200 kV TEM (Talos F200X, FEI) equipped with Lorentz mode.

*First-principles calculations:* The first-principles calculations were performed within the framework of density functional theory (DFT) as implemented in the Vienna ab-initio simulation package (VASP)[36]. The interaction between valence and core electrons was treated by the projector augmented wave (PAW) method[37]. The spin-polarized generalized gradient approximation (GGA) with the Perdew–Burke–Ernzerhof (PBE) version was used to describe exchange-correlation energy[38]. Since the GGA functional fails to treat partially occupied 3d electrons of transition metal elements, we employed the GGA+U method with an effective $U$ = 3 eV for Fe element as reported in the previous studies[39, 40]. The interlayer van der Waals (vdW) interactions of bulk and bilayer $Fe_3GaTe_{2-x}$ (x=0.25) were described with the DFT-D3 correction in Grimme's scheme[41]. A kinetic cutoff energy of 450 eV was used for the plane-wave expansion. The $k$-point sampling in the Brillouin zone was implemented using the Monkhorst–Pack scheme[42] with a grid of 3×12×1 for pristine and oxidized bilayer $Fe_3GaTe_{2-x}$ and 2×8×2 for pristine bulk $Fe_3GaTe_{2-x}$, respectively. All the structures were fully relaxed until the force acting on each atom was less than 0.01 eV·Å$^{-1}$. The Bader charge analysis[43] was carried out to examine the charge transfers among O dopants, Fe, Ga, and Te atoms in the oxidized bilayer $Fe_3GaTe_{2-x}$.

We used the chirality-dependent total energy difference approach to obtain the DMI strength[44, 45]. The DMI energy between normalized spins restricted to the nearest neighbors can be determined by:

$$E_{DMI} = \sum_{\langle i,j \rangle} \boldsymbol{d}_{ij} \cdot (\boldsymbol{S}_j \times \boldsymbol{S}_j) \quad (1)$$

where $\mathbf{d}_{ij}$ is summed by considering two types of pairs, those inside a given layer *L*, and interlayer pairs between a layer *L* and layers above or below. From the Moriya symmetry rule[46], the DMI vector for the layer *L* can be written as:

$$\mathbf{d}_{ij}^{L} = d^{L}(\hat{\mathbf{z}} \times \hat{\mathbf{u}}_{ij}) \quad (2)$$

where $\hat{\mathbf{z}}$ and $\hat{\mathbf{u}}_{ij}$ are unit vectors pointing along *z* and from site *i* to site *j*, respectively. The total DMI strength, $d^{tot}$ is derived by identifying the difference between the DFT energies $E_{CW}$ and $E_{ACW}$ for opposite chirality spin configuration (**Figure S18, Supporting Information**) with the corresponding energy differences calculated as follows:

$$d^{tot} = (E_{CW} - E_{ACW})/m \quad (3)$$

where the number *m* depends on the length of cycloidal unit cell. For example, *m* is 12 for the cycloid unit cell with *n* = 4 (*n* is primitive cell number along the cycloid direction), as shown in **Figure S18, Supporting Information**. It needs to be emphasized that $d^{tot}$ is the DMI strength in a single atomic layer and producing an equivalent effect. The global effect on the bilayer or multilayer can also be expressed by the micromagnetic energy per volume unit of the magnetic film.

*Statistical Analysis:* 1) The intensities of Figure 2f and Figure S1 (Supporting Information) are normalized. 2) The error bars in Figures 2f,g, 4c,d and Figures S3e, S5f, S6d,e, S15, S17 (Supporting Information) are presented by mean ± standard deviation (sd). 3) The sample sizes (N) are presented in the caption of each figure. 4) Microsoft Office Excel, Origin and MATLAB are used for statistical analysis.

## Supporting Information

Supporting Information is available from the Wiley Online Library or from the author.

## Acknowledgements

H.C. designed the project. G.Z. and W.J. grew the single crystals. G.Z. fabricated the Hall devices. G.Z., H.W. and L.Y. did the characterization and measurements. X.W. drew the skyrmion phase diagrams. H.S, Q.L, and J.Z did the theoretical calculations. H.C., G.Z., W.Z. and L.L. analyzed the results. G.Z. and H.C. wrote the paper. This work was supported by the National Key Research and Development Program of China (2022YFE0134600), National Natural Science Foundation of China (52272152, 61674063 and 62074061), Shenzhen Science and Technology Innovation Committee (JCYJ20210324142010030), Natural Science Foundation of Hubei Province, China (2022CFA031), and Interdisciplinary Research Program of Huazhong University of Science and Technology (5003110122). The Analytical and Testing Center in Huazhong University of Science and Technology for EPMA and XPS tests are acknowledged.

## Conflict of Interest

The authors declare that they have no competing interests.

## Data Availability Statement

The data that support the findings of this study are available from the corresponding author upon reasonable request.

## References

[1] H. Jani, J. C. Lin, J. Chen, J. Harrison, F. Maccherozzi, J. Schad, S. Prakash, C. B. Eom, A. Ariando, T. Venkatesan, P. G. Radaelli, *Nature* **2021**, 590, 74.

[2] Wanjun Jiang, Pramey Upadhyaya, Wei Zhang, Guoqiang Yu, M. Benjamin Jungfleisch, Frank Y. Fradin, John E. Pearson, Yaroslav Tserkovnyak, Kang L. Wang, Olle Heinonen, Suzanne G. E. te Velthuis, A. Hoffmann, *Science* **2015**, 349, 283.

[3] A. Fert, N. Reyren, V. Cros, *Nat. Rev. Mater.* **2017**, 2, 17031.

[4] S. Mühlbauer, B. Binz, F. Jonietz, C. Pfleiderer, A. Rosch, A. Neubauer, R. Georgii, P. Böni, *Science* **2009**, 323, 915.

[5] X. Z. Yu, Y. Onose, N. Kanazawa, J. H. Park, J. H. Han, Y. Matsui, N. Nagaosa, Y. Tokura, *Nature* **2010**, 465, 901.

[6] Hongrui Zhang, David Raftrey, Ying-Ting Chan, Yu-Tsun Shao, Rui Chen, Xiang Chen, Xiaoxi Huang, Jonathan T. Reichanadter, Kaichen Dong, Sandhya Susarla, Lucas Caretta, Zhen Chen, Jie Yao, Peter Fischer, Jeffrey B. Neaton, Weida Wu, David A. Muller, Robert J. Birgeneau, R. Ramesh, *Sci. Adv.* **2022**, 8, eabm7103.

[7] Takashi Kurumaji, Taro Nakajima, Max Hirschberger, Akiko Kikkawa, Yuichi Yamasaki, Hajime Sagayama, Hironori Nakao, Yasujiro Taguchi, Taka-hisa Arima, Y. Tokura, *Science* **2019**, 365, 914.

[8] Elizabeth Skoropata, John Nichols, Jong Mok Ok, Rajesh V. Chopdekar, Eun Sang Choi, Ankur Rastogi, Changhee Sohn, Xiang Gao, Sangmoon Yoon, Thomas Farmer, Ryan D. Desautels, Yongseong Choi, Daniel Haskel, John W. Freeland, Satoshi Okamoto, Matthew Brahlek, H. N. Lee, *Sci. Adv.* **2020**, 6, eaaz3902.

[9] Q. Shao, Y. Liu, G. Yu, S. K. Kim, X. Che, C. Tang, Q. L. He, Y. Tserkovnyak, J. Shi, K. L. Wang, *Nat. Electron.* **2019**, 2, 182.

[10] Y. Wu, S. Zhang, J. Zhang, W. Wang, Y. L. Zhu, J. Hu, G. Yin, K. Wong, C. Fang, C. Wan, X. Han, Q. Shao, T. Taniguchi, K. Watanabe, J. Zang, Z. Mao, X. Zhang, K. L. Wang, *Nat. Commun.* **2020**, 11, 3860.

[11] A. Soumyanarayanan, M. Raju, A. L. Gonzalez Oyarce, A. K. C. Tan, M. Y. Im, A. P. Petrovic, P. Ho, K. H. Khoo, M. Tran, C. K. Gan, F. Ernult, C. Panagopoulos, *Nat. Mater.* **2017**, 16, 898.

[12] X. Zhang, S. C. Ambhire, Q. Lu, W. Niu, J. Cook, J. S. Jiang, D. Hong, L. Alahmed, L. He, R. Zhang, Y. Xu, S. S. Zhang, P. Li, G. Bian, *ACS Nano* **2021**, 15, 15710.

[13] Y. Wu, B. Francisco, Z. Chen, W. Wang, Y. Zhang, C. Wan, X. Han, H. Chi, Y.

Hou, A. Lodesani, G. Yin, K. Liu, Y. T. Cui, K. L. Wang, J. S. Moodera, *Adv. Mater.* **2022**, 34, e2110583.
[14] D. Liang, J. P. DeGrave, M. J. Stolt, Y. Tokura, S. Jin, *Nat. Commun.* **2015**, 6, 8217.
[15] T. Schulz, R. Ritz, A. Bauer, M. Halder, M. Wagner, C. Franz, C. Pfleiderer, K. Everschor, M. Garst, A. Rosch, *Nat. Phys.* **2012**, 8, 301.
[16] Z. S. Lim, C. Li, Z. Huang, X. Chi, J. Zhou, S. Zeng, G. J. Omar, Y. P. Feng, A. Rusydi, S. J. Pennycook, T. Venkatesan, A. Ariando, *Small* **2020**, 16, e2004683.
[17] R. Ritz, M. Halder, C. Franz, A. Bauer, M. Wagner, R. Bamler, A. Rosch, C. Pfleiderer, *Phys. Rev. B* **2013**, 87, 134424.
[18] S. X. Huang, C. L. Chien, *Phys. Rev. Lett.* **2012**, 108, 267201.
[19] H. Li, B. Ding, J. Chen, Z. Li, Z. Hou, E. Liu, H. Zhang, X. Xi, G. Wu, W. Wang, *Applied Physics Letters* **2019**, 114.
[20] G. Zhang, F. Guo, H. Wu, X. Wen, L. Yang, W. Jin, W. Zhang, H. Chang, *Nat. Commun.* **2022**, 13, 5067.
[21] G. Kimbell, C. Kim, W. Wu, M. Cuoco, J. W. A. Robinson, *Commun Mater* **2022**, 3, 19.
[22] L. Tai, B. Dai, J. Li, H. Huang, S. K. Chong, K. L. Wong, H. Zhang, P. Zhang, P. Deng, C. Eckberg, G. Qiu, H. He, D. Wu, S. Xu, A. Davydov, R. Wu, K. L. Wang, *ACS Nano* **2022**, 16, 17336.
[23] J. H. Jeon, H. R. Na, H. Kim, S. Lee, S. Song, J. Kim, S. Park, J. Kim, H. Noh, G. Kim, S. K. Jerng, S. H. Chun, *ACS Nano* **2022**, 16, 8974.
[24] T.-E. Park, L. Peng, J. Liang, A. Hallal, F. S. Yasin, X. Zhang, K. M. Song, S. J. Kim, K. Kim, M. Weigand, G. Schütz, S. Finizio, J. Raabe, K. Garcia, J. Xia, Y. Zhou, M. Ezawa, X. Liu, J. Chang, H. C. Koo, Y. D. Kim, M. Chshiev, A. Fert, H. Yang, X. Yu, S. Woo, *Phys. Rev. B* **2021**, 103, 104410.
[25] L. Peng, F. S. Yasin, T. E. Park, S. J. Kim, X. Zhang, T. Nagai, K. Kimoto, S. Woo, X. Yu, *Adv. Funct. Mater.* **2021**, 31, 2103583.
[26] W. Wang, Y. F. Zhao, F. Wang, M. W. Daniels, C. Z. Chang, J. Zang, D. Xiao, W. Wu, *Nano Lett.* **2021**, 21, 1108.
[27] M. V. Sapozhnikov, N. S. Gusev, S. A. Gusev, D. A. Tatarskiy, Y. V. Petrov, A. G. Temiryazev, A. A. Fraerman, *Phys. Rev. B* **2021**, 103, 054429.
[28] J. Zang, M. Mostovoy, J. H. Han, N. Nagaosa, *Phys. Rev. Lett.* **2011**, 107, 136804.
[29] N. Nagaosa, Y. Tokura, *Nat. Nanotechnol.* **2013**, 8, 899.
[30] S. Woo, K. Litzius, B. Kruger, M. Y. Im, L. Caretta, K. Richter, M. Mann, A. Krone, R. M. Reeve, M. Weigand, P. Agrawal, I. Lemesh, M. A. Mawass, P. Fischer, M. Klaui, G. S. Beach, *Nat. Mater.* **2016**, 15, 501.
[31] S. Woo, K. M. Song, X. Zhang, Y. Zhou, M. Ezawa, X. Liu, S. Finizio, J. Raabe, N. J. Lee, S. I. Kim, S. Y. Park, Y. Kim, J. Y. Kim, D. Lee, O. Lee, J. W. Choi, B. C. Min, H. C. Koo, J. Chang, *Nat. Commun.* **2018**, 9, 959.
[32] G. Yu, P. Upadhyaya, Q. Shao, H. Wu, G. Yin, X. Li, C. He, W. Jiang, X. Han, P. K. Amiri, K. L. Wang, *Nano Lett.* **2017**, 17, 261.
[33] Gong Chen, Arantzazu Mascaraque, Hongying Jia, Bernd Zimmermann, MacCallum Robertson, Roberto Lo Conte, Markus Hoffmann, Miguel Angel González Barrio, Haifeng Ding, Roland Wiesendanger, Enrique G. Michel, Stefan Blügel,

Andreas K. Schmid, K. Liu, *Sci. Adv.* **2020**, 6, eaba4924.
[34] M. Arora, J. M. Shaw, H. T. Nembach, *Phys. Rev. B* **2020**, 101, 054421.
[35] A. Belabbes, G. Bihlmayer, S. Blugel, A. Manchon, *Sci. Rep.* **2016**, 6, 24634.
[36] G. Kresse, J. Furthmu¨ller, *Phys. Rev. B* **1996**, 54, 11169.
[37] P. E. Blochl, O. Jepsen, O. K. Andersen, *Phys. Rev. B* **1994**, 49, 16223.
[38] G. Kresse, J. Hafner, *Phys. Rev. B* **1993**, 47, 558.
[39] M. C. Payne, M. P. Teter, D. C. Allan, T. A. Arias, J. D. Joannopoulos, *Rev. Mod. Phys.* **1992**, 64, 1045.
[40] G. Kresse, D. Joubert, *Phys. Rev. B* **1998**, 59, 1758.
[41] S. Grimme, *J. Comput. Chem.* **2006**, 27, 1787.
[42] H. J. Xiang, E. J. Kan, S.-H. Wei, M. H. Whangbo, X. G. Gong, *Phys. Rev. B* **2011**, 84, 224429.
[43] W. Tang, E. Sanville, G. Henkelman, *J Phys Condens Matter* **2009**, 21, 084204.
[44] J. H. Yang, Z. L. Li, X. Z. Lu, M. H. Whangbo, S. H. Wei, X. G. Gong, H. J. Xiang, *Phys. Rev. Lett.* **2012**, 109, 107203.
[45] H. Yang, A. Thiaville, S. Rohart, A. Fert, M. Chshiev, *Phys. Rev. Lett.* **2015**, 115, 267210.
[46] T. Moriya, *Phys. Rev.* **1960**, 120, 91.

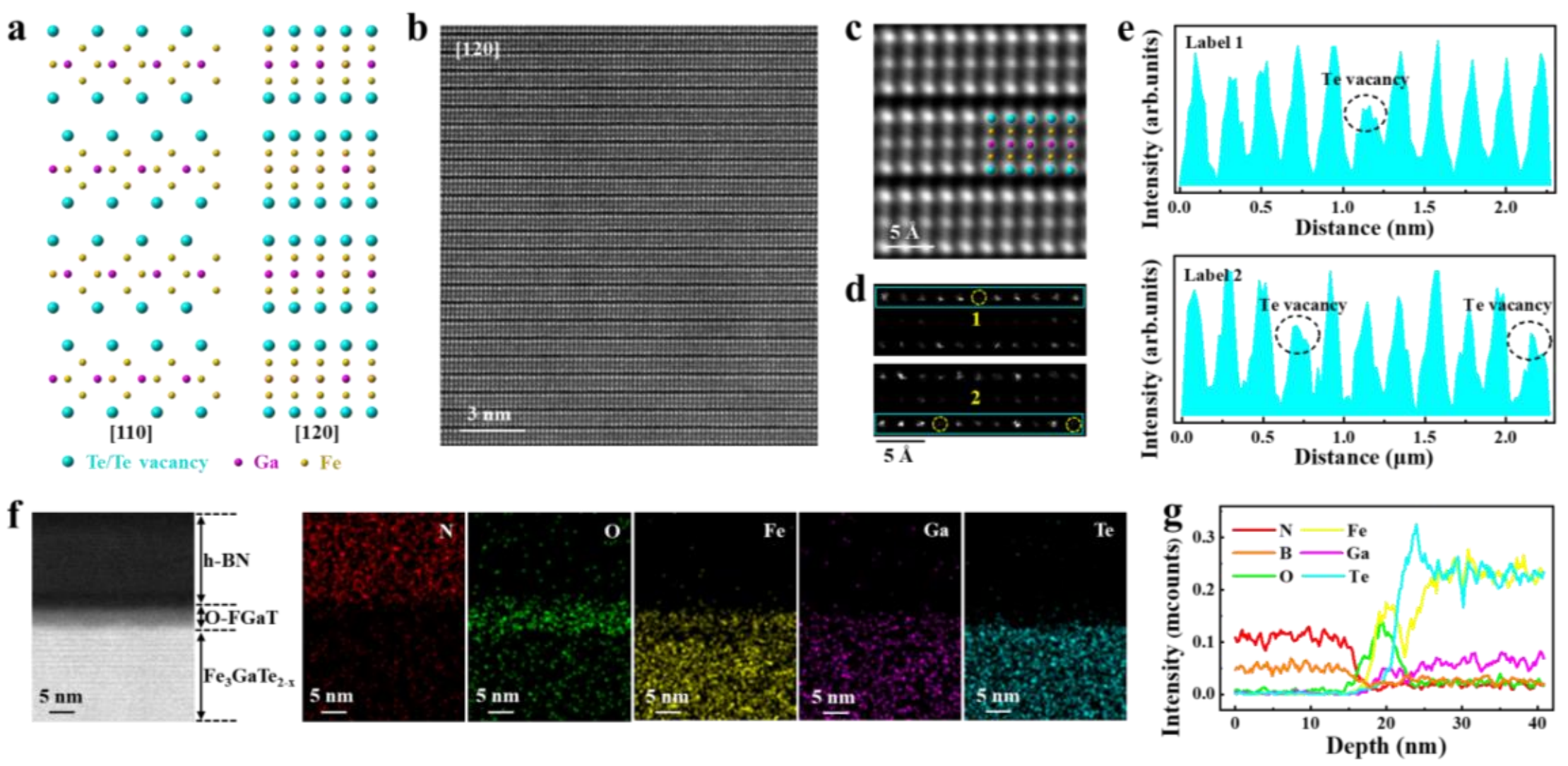

**Figure 1. Crystal structure and TEM characterizations of 2D O-FGaT/FGaT heterostructure from natural oxidization.** a) Structural models of vdW $Fe_3GaTe_{2-x}$ along the [110] and [120] directions. b) Cross-sectional HAADF-STEM image of vdW $Fe_3GaTe_{2-x}$ crystal along the [120] direction. c) Atomic-resolution HAADF-STEM image and corresponded arrangement of the stacking structure of Te, Fe, and Ga atoms. d,e) Observation of Te vacancies and corresponded intensity profiles. The visibility and contrast between atoms were enhanced for identifying the Te vacancy. f) Cross-sectional HAADF image and corresponded elemental mapping of an h-BN capped O-FGaT/FGaT. Note that few diluted scattered oxygen distribution imaged within $Fe_3GaTe_{2-x}$ comes from the temporary atmospheric exposure during the sample preparation by FIB. g) Elemental line distribution along the depth direction.

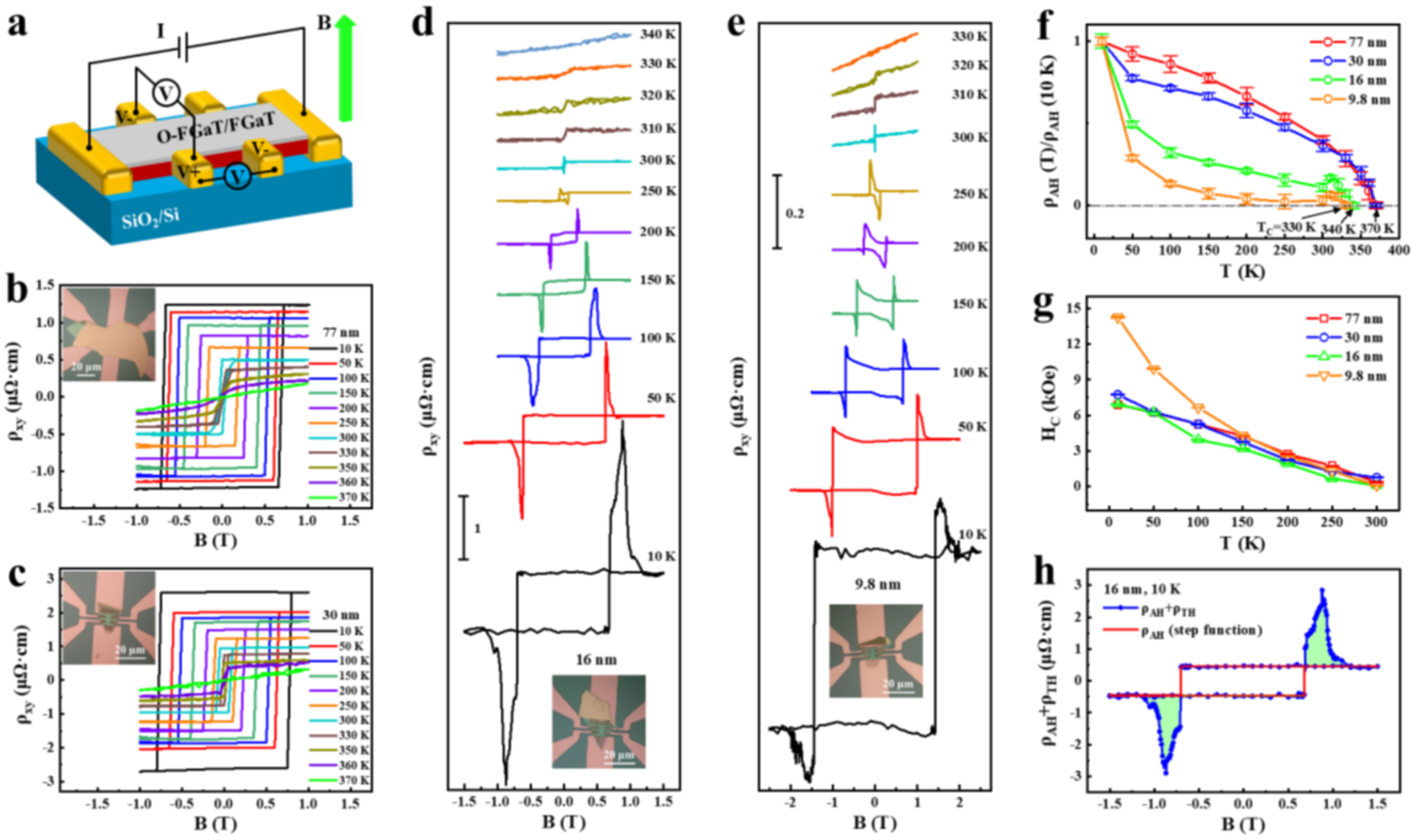

**Figure 2. Thickness-dependent magnetotransport of 2D room-temperature ferromagnetic O-FGaT/FGaT heterostructures.** a) Schematic illustration and measurement geometry of Hall devices. b-e) $\rho_{xy}$-B curves for four 2D O-FGaT/FGaT with different thickness at varying temperatures. Insets show optical images of each Hall device. f,g) Temperature-dependent normalized $\rho_{AH}$ and $H_C$ extracted from (b-e). Error bars sd., N=25 for $\rho_{AH}$ and N=3 for $H_C$. The temperature at zero normalized $\rho_{AH}$ is determined to be the $T_C$. h) Diagram of extracting the THE contribution in a typical 16 nm 2D O-FGaT/FGaT at 10 K. Contribution from AHE (red solid line) and THE term (light green area) is marked, where the AHE contribution is fitted by a step function.

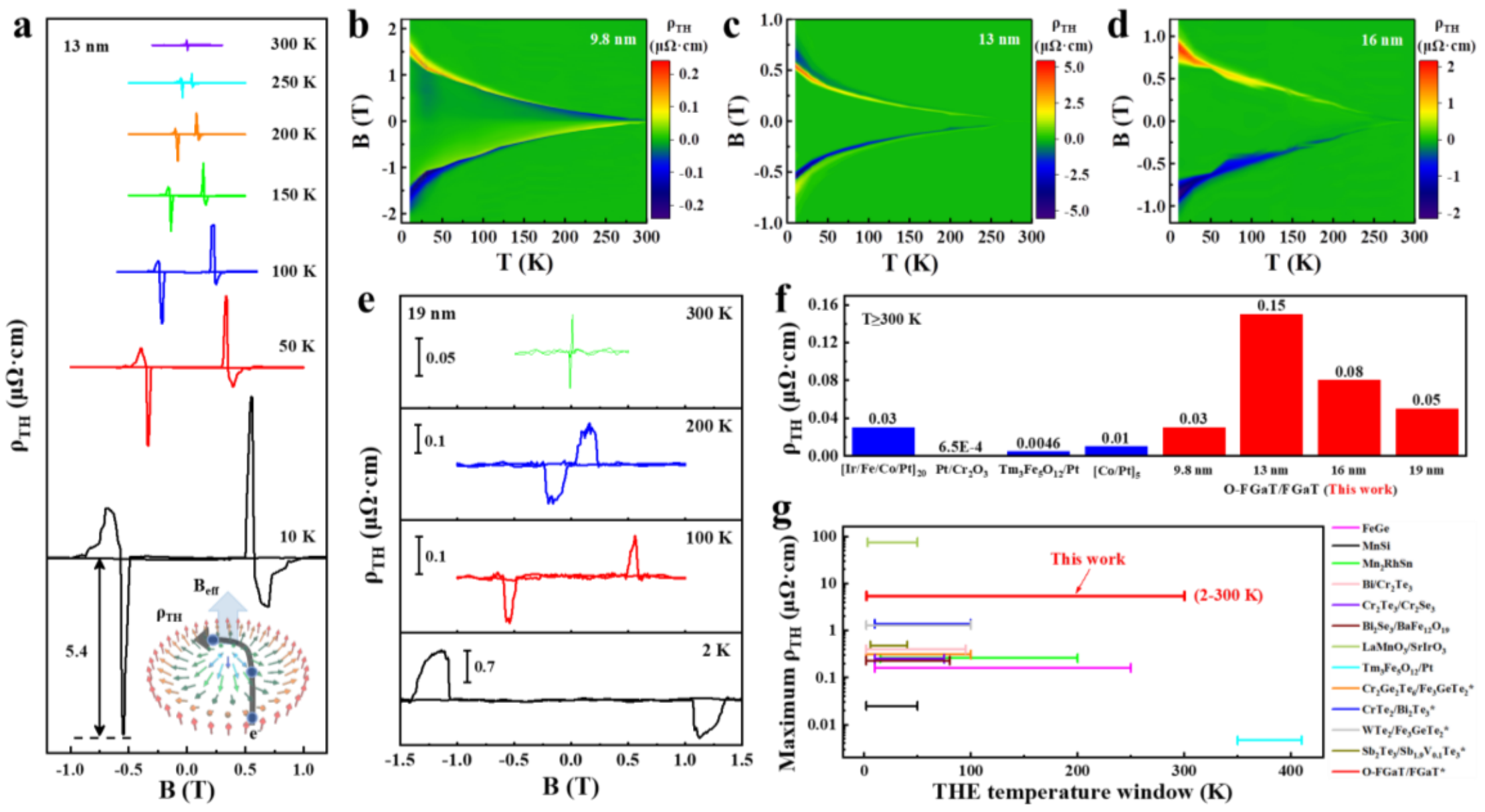

**Figure 3. Giant 2D THE in sub-20 and sub-10 nm 2D room-temperature ferromagnetic O-FGaT/FGaT heterostructures.** a) $\rho_{TH}$-B curves for a 13 nm O-FGaT/FGaT at different temperatures. Inset, the scheme of skyrmion-based THE. b-d) Skyrmion phase diagrams from the THE as a function of temperature and magnetic field for 9.8, 13 and 16 nm O-FGaT/FGaT. The color bar indicates the value of $\rho_{TH}$. Interpolation is performed between each experimental data point. e) $\rho_{TH}$-B curves for a 19 nm O-FGaT/FGaT at different temperatures ranging from 2 to 300 K. f) Comparison of the high-temperature (≥300 K) $\rho_{TH}$ with that of other 2D skyrmion systems. g) Comparison of the maximum $\rho_{TH}$ and THE temperature window with that of other 2D skyrmion systems. See reported data and references in **Table S2, Supporting Information**.

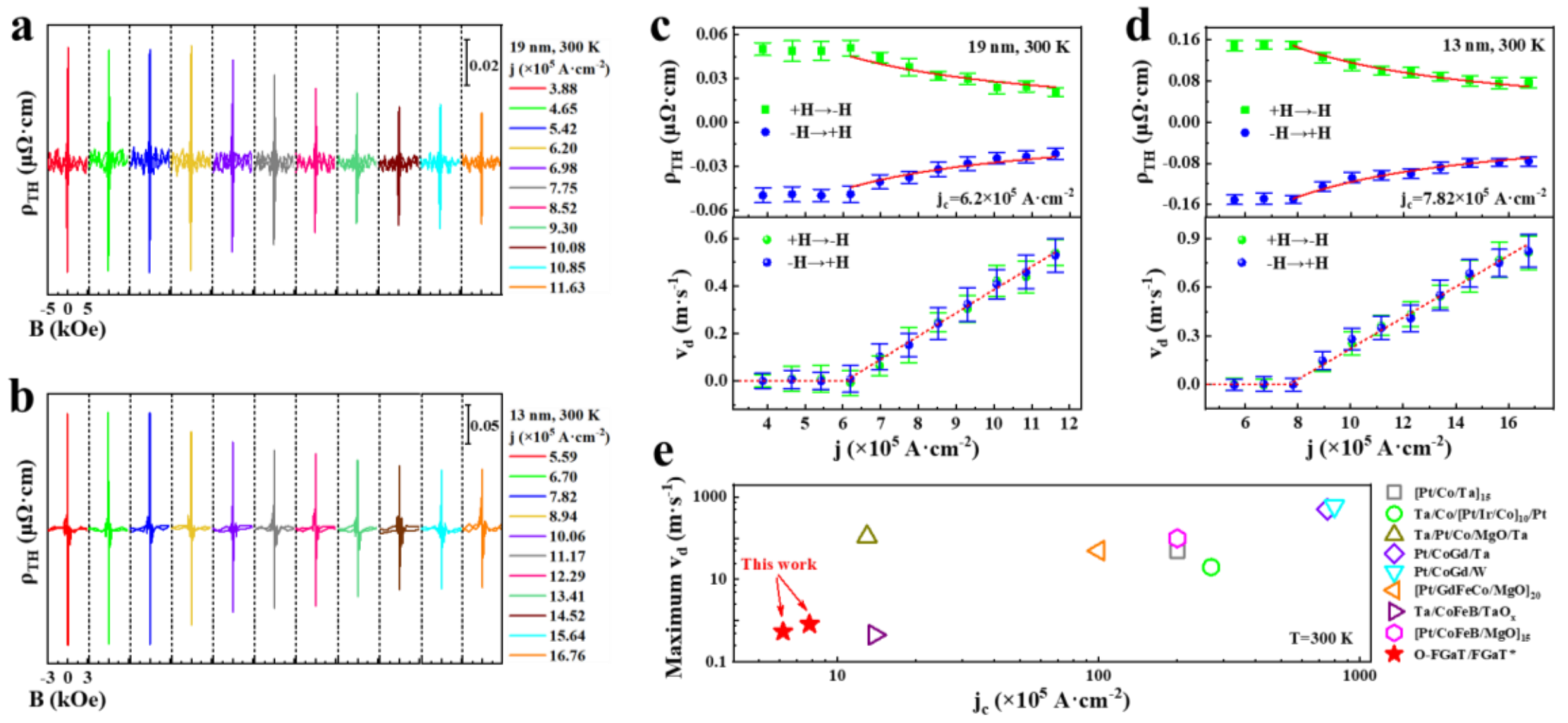

**Figure 4. Room-temperature current-controlled 2D THE by low critical current density ($j_c$) in 2D O-FGaT/FGaT heterostructures.** a,b) $\rho_{TH}$-B curves of the 19 and 13 nm O-FGaT/FGaT at 300 K under increasing current densities. c,d) Current density (j) dependence of the $\rho_{TH}$ and $v_d$ for 19 and 13 nm O-FGaT/FGaT. The red lines are fitted curves. Error bars sd., N=25. e) Room-temperature $j_c$ and maximum $v_d$ comparison for various 2D skyrmion systems with vdW structures (solid symbols) and non-vdW structures (open symbols). See reported data and references in **Table S3, Supporting Information**.

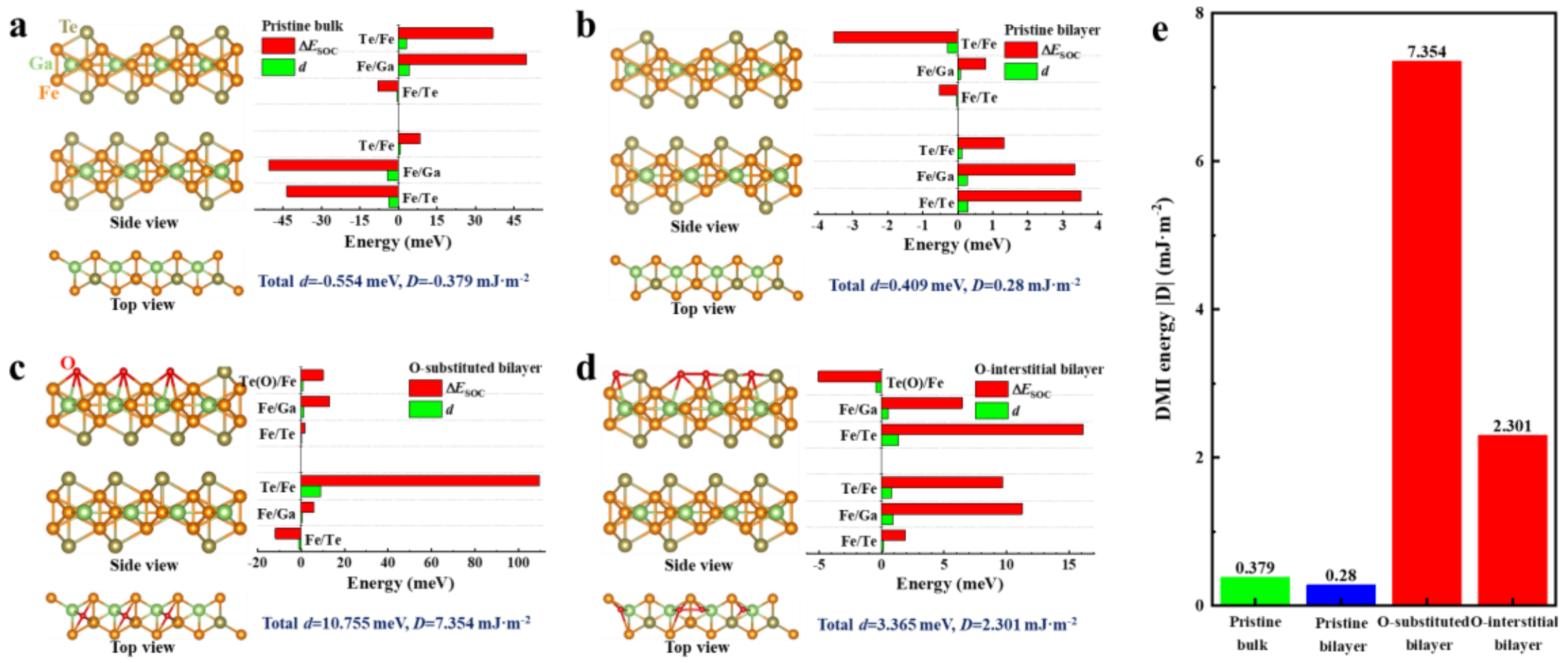

**Figure 5. First-principles calculations of DMI in pristine and oxidized $Fe_3GaTe_{2-x}$.** a) Atomic model, layer-resolved SOC energy difference ($\Delta E_{SOC}$), and DMI parameter ($d$) in pristine bulk $Fe_3GaTe_{2-x}$. b-d) Atomic models, layer-resolved $\Delta E_{SOC}$, and $d$ in pristine (b), O-substituted (c) and O-interstitial (d) bilayer $Fe_3GaTe_{2-x}$. The inserted values below each case show total microscopic and micromagnetic DMI parameters ($d$ and $D$). e) Comparison of DMI energies (|D|) for bilayer and bulk $Fe_3GaTe_{2-x}$ with and/or without oxidization.

# Supporting Information for

# Giant 2D Skyrmion Topological Hall Effect with Ultrawide Temperature Window and Low-Current Manipulation in 2D Room-Temperature Ferromagnetic Crystals

*Gaojie Zhang, Qingyuan Luo, Xiaokun Wen, Hao Wu*[*]*, Li Yang, Wen Jin, Luji Li, Jia Zhang, Wenfeng Zhang, Haibo Shu*[*]*, Haixin Chang*[*]

[*]Corresponding authors. E-mail: hxchang@hust.edu.cn, h_wu@hust.edu.cn, shuhaibo@cjlu.edu.cn

**This file includes:**

Notes S1-S5

Figures S1-S19

Tables S1-S5

**Notes S1-S5:**

**Note S1.** XPS results under different etching time for surface oxidization analysis

**Note S2.** Strong intrinsic ferromagnetism in the bulk $Fe_3GaTe_{2-x}$ crystal and 2D $Fe_3GaTe_{2-x}$ nanosheet

**Note S3.** Discussion of excluding the artifact "THE" signals

**Note S4.** Additional analysis of THE and resulted THE-derived skyrmion sizes

**Note S5.** Discussion of Joule heating effect in current-controlled THE at room temperature

**Figures S1-S19:**

**Figure S1.** Crystal photograph and XRD pattern of the bulk $Fe_3GaTe_{2-x}$ crystals.

**Figure S2.** TEM characterization of a 2D $Fe_3GaTe_{2-x}$ nanosheet along the [001] direction.

**Figure S3.** Quantitative analysis of Fe, Ga and Te content in three pristine $Fe_3GaTe_{2-x}$ single-sheet nanosheets by EPMA.

**Figure S4.** XPS analysis on an O-FGaT/FGaT surface under different etching time.

**Figure S5.** Above-room-temperature strong ferromagnetism in pristine vdW bulk $Fe_3GaTe_{2-x}$ crystals.

**Figure S6.** Magneto-transport measurement of a non-oxidized pristine 14 nm $Fe_3GaTe_{2-x}$ nanosheet.

**Figure S7.** AFM images and the corresponded profile height of six as-tested Hall devices based on 2D O-FGaT/FGaT heterostructures.

**Figure S8.** Magneto-transport measurement of a 13 nm 2D O-FGaT/FGaT heterostructure.

**Figure S9.** Natural total oxidation of the 2D $Fe_3GaTe_{2-x}$ nanosheets in the air for 48 h.

**Figure S10.** Observation of Néel-type skyrmions in 2D O-FGaT/FGaT by Lorentz-TEM with the perpendicular magnetic field at 300 K.

**Figure S11.** Magnetic-field-driven evolution from stripe domains to skyrmions in a thin 2D O-FGaT/FGaT by Lorentz-TEM with perpendicular magnetic field at 300 K.

**Figure S12.** Lorentz-TEM images under different perpendicular magnetic field in a pristine non-oxidized thin 2D $Fe_3GaTe_{2-x}$ nanosheet at 300 K under α=29°, d=-2 mm.

**Figure S13.** Extracting the THE signals in 2D O-FGaT/FGaT by step function.

**Figure S14.** Temperature-dependent $\rho_{TH}$-B curves for two 2D O-FGaT/FGaT heterostructures with different thickness.

**Figure S15.** Magnetic field dependence of the skyrmion density and $\rho_{TH}$ at room temperature in 2D O-FGaT/FGaT.

**Figure S16.** Comparison of THE tests-derived minimum skyrmion size ($n_{sk}^{-1/2}$) in various 2D skyrmion systems.

**Figure S17.** $\rho_{xx}$ and saturated $\rho_{AH}$ as a function of current densities in 2D O-FGaT/FGaT at room temperature.

**Figure S18.** Theoretical model for calculating DMI in $Fe_3GaTe_{2-x}$.

**Figure S19.** The density of states (DOSs) comparison of total, Fe-3d, Ga-4p, and Te-5p in oxidized bilayer $Fe_3GaTe_{2-x}$ with that of pristine bilayer $Fe_3GaTe_{2-x}$.

**Tables S1-S5:**

**Table S1.** Magneto-transport parameters for THE calculation in 2D O-FGaT/FGaT heterostructures with different thickness.

**Table S2.** Comparison of the THE temperature window and maximum $\rho_{TH}$ in various 2D skyrmion systems from the literatures.

**Table S3.** Comparison of critical current density ($j_c$) and maximum drift velocity ($v_d$) in various room-temperature 2D skyrmion systems from literatures.

**Table S4.** The binding energies ($E_b$) for incorporating oxygen into $Fe_3GaTe_{2-x}$ and $Fe_3GaTe_2$ crystals.

**Table S5.** Average Bader charges (Q) of single atom in pristine, O-substituted, and O-interstitial bilayer $Fe_3GaTe_{2-x}$.

**Note S1. XPS results under different etching time for surface oxidation analysis**

In the initial state without etching, the Fe 2p spectra is decomposed into Fe $2p_{3/2}$ and Fe $2p_{1/2}$ (**Figure S4a**). Specifically, two peaks at 706.7 and 719. 8 eV come from the Fe(0). The other two peaks at 710.8 and 724.4 eV are ascribed to the Fe(III). Meanwhile, the Ga 2p spectra is decomposed into Ga $2p_{3/2}$ and Ga $2p_{1/2}$ peaks with binding energies of 1117.8 and 1144.7 eV, corresponding to the natively-oxidized Ga(II) (**Figure S4b**). And then, the Te spectra can be decomposed into Te $3d_{5/2}$ and Te $3d_{3/2}$. Two peaks at 572.8 and 583.2 eV correspond to the Te (II). Other four peaks at 576.3, 586.7, 573.9, and 584.4 eV all originate from the oxidized Te (**Figure S4c**). Moreover, the O 1s spectra are decomposed into three peaks including 530.5, 531.6, and 532.9 eV, corresponding to the substitutional oxygen, interstitial oxygen and adsorbed oxygen, respectively (**Figure S4d**)[1].

With the increase of etching time, the oxidation peaks of Te 3d gradually decrease and eventually disappear. At the same time, note that the emergence of Fe(II), Ga(II) peaks and the increase of Fe(0), Fe(II), Ga(II) peaks are attributed to the reduction effect of $Ar^+$ during the etching process, which can also be seen in other reports[2, 3]. Together with cross-sectional TEM imaging and elemental mapping, all these results indicate the presence of an ultrathin O-FGaT layer on the surface of the $Fe_3GaTe_{2-x}$ crystal.

**Note S2. Strong intrinsic ferromagnetism in the bulk $Fe_3GaTe_{2-x}$ crystal and 2D $Fe_3GaTe_{2-x}$ nanosheet**

The temperature dependent magnetization (M-T) curves under zero-field-cooling and field-cooling (ZFC-FC) regime of bulk $Fe_3GaTe_{2-x}$ crystal exhibit a typically ferromagnetic feature and an above-room-temperature $T_C$ (~358 K) (**Figure S5a,b**), higher than most known 2D vdW ferromagnets[4-7]. Moreover, the magnetization of out-of-plane ZFC-FC curve is larger than that of in-plane ZFC-FC curve, demonstrating the PMA in bulk $Fe_3GaTe_{2-x}$ crystal. A same conclusion can also be obtained from the out-of-plane and in-plane magnetized M-B curves with significantly different shapes (**Figure S5c,d**). The PMA energy density ($K_u$) is determined by the following formula[8]:

$$K_u=\frac{B_{sat}M_{sat}}{2} \quad (1)$$

where $B_{sat}$ is the saturation field of hard axis, $M_{sat}$ is the saturation magnetization. Among them, the $B_{sat}$ of bulk $Fe_3GaTe_{2-x}$ crystal is determined by VSM test under 300 K and in-plane magnetic field from -9 to 9 T (**Figure S5e**). The $M_{sat}$ and $H_C$ of bulk $Fe_3GaTe_{2-x}$ crystal are measured by VSM tests under different temperatures (**Figure S5f**). Therefore, the $K_u$ of bulk $Fe_3GaTe_{2-x}$ crystal is calculated as ~$4.65\times10^5$ J/$m^3$ at 300 K, consistent with the previous report[9]. Such large room-temperature $K_u$ is one order of magnitude larger than known vdW ferromagnets (*e.g.* $CrTe_2$[10]) and is better than non-vdW ferromagnetic films (*e.g.* CoFeB[8]). The bulk $Fe_3GaTe_{2-x}$ crystal is a hard ferromagnet at 10 K ($H_C$, ~830 Oe) and turn into a soft ferromagnet at 300 K ($H_C$, ~200 Oe). In addition, the $M_{sat}$ of bulk $Fe_3GaTe_{2-x}$ crystal is ~55.1 emu/g at 10 K and remain ~36.3 emu/g at 300 K, ~2.9 times that in 2D vdW ferromagnet $CrTe_2$ at room temperature[7].

To further study the 2D ferromagnetism, the magneto-transport measurement is performed on a non-oxidized 14 nm $Fe_3GaTe_{2-x}$ nanosheet (**Figure S6a**). The typical metallic characteristic is observed from the temperature-dependent longitudinal resistivity ($\rho_{xx}$-T) curve, where the $\rho_{xx}$ (300 K)=$3.24\times10^2$ μΩ·cm (**Figure S6b**). As shown in **Figure S6c**, the AHE exists when the temperature is below $T_C$ ~350 K and the square hysteresis loop with nearly vertical magnetization flipping persists at ~320 K, demonstrating the coexistence of long-range ferromagnetism and large PMA at above room temperature in 2D $Fe_3GaTe_{2-x}$ nanosheet. Compared with 2D O-FGaT/FGaT heterostructures, the 14 nm $Fe_3GaTe_{2-x}$ nanosheet shows no THE in the $\rho_{xy}$-B curves at all temperatures. Further, we carefully record the $\rho_{AH}$ and $H_C$ as a function of temperature, thereby implying the influence of thermal fluctuation on 2D ferromagnetism (**Figure S6d,e**). Unlike $H_C$, which decreases with increasing temperature, the $\rho_{AH}$ is almost constant at first, and then gradually decreases after the temperature exceeds 150 K.

**Note S3. Discussion of excluding the artifact "THE" signals**

Recent criticisms about some THE may come from the artificial multiple conduction channels, since most hump and dip features can indeed be superposed by two-component AHE with opposite signs and different $H_C$[11-15]. Identifying this concern requires understanding the rationale behind it. An artifact "THE" mainly happens in heterostructures with parallel multi-conduction channels, or in inhomogeneous ferromagnets[16]. Fortunately, Seung-Hyun Chun et al.[17] and Kang L. Wang et al.[18] reported some guidance methods for distinguishing the artifact “THE” and real THE.

For the concern of parallel multi-conduction channels, one example is a $(Bi, Sb)_2Te_3$/(V, Bi, Sb)$_2Te_3$ heterostructure which contains both surface and bulk ferromagnetism, forming two-component AHE with opposite signs and different $H_C$[12]. To discuss this case, we let a 2D $Fe_3GaTe_{2-x}$ nanosheet oxidize naturally in the air for 48 h and record the change of $\rho_{xx}$ at 300 K (**Figure S9a**). Note that oxidation for 48 h is sufficient to completely oxidize this $Fe_3GaTe_{2-x}$ nanosheet, forming an ~8 nm O-FGaT layer confirmed by AFM and cross-sectional TEM (**Figure S9b,c**). The room-temperature $\rho_{xx}$ of this 8 nm O-FGaT layer reaches to $1.23\times10^7$ μΩ·cm, which is ~$10^5$ times that in non-oxidized 2D $Fe_3GaTe_{2-x}$ nanosheet (**Figure S6b**). Therefore, we believe that the current totally flow in the $Fe_3GaTe_{2-x}$, as the resistivity of O-FGaT layer is too large to contribute to the conduction. A similar method of excluding artifact “THE” has been endorsed in another report[17].

For the concern of inhomogeneous ferromagnets, a ferromagnet with inhomogeneous thickness (*e.g.* $SrRuO_3$[14, 15]) or containing hidden ferromagnetic phase (*e.g.* $MnBi_2Te_4$ films containing $MnTe_2$[18]) sometimes leads to two spatially separated ferromagnetic regions and present two-component AHE with opposite signs and different $H_C$[13]. The difference in the temperature dependence of such two-component AHE results in the polarity change of the total AHE in a narrow temperature range, where an artifact “THE” occurs[11, 14]. In contrast, temperature-dependent real THE with AHE reveals no such polarity change of AHE, as real THE only occurs around the spin-flipping region of the AHE and should not affect the polarity of the AHE[18]. In this work, no temperature- or thickness-induced polarity change of AHE are observed in all 2D O-FGaT/FGaT heterostructures (**main text Figure 2d,e** and **Figure S8c**), and robust 2D THE exists in a wide temperature window ranging from 2 to 300 K (**main text Figure 3e**). Also, as a controlled sample, the non-oxidized 14 nm $Fe_3GaTe_{2-x}$ nanosheet shows neither the temperature-induced polarity change of AHE nor the THE (**Figure S6c**), and the selected-area electron diffraction (SAED) result for non-oxidized $Fe_3GaTe_{2-x}$ shows a set of clean diffraction spots for single crystals without hidden ferromagnetic phase (**Figure S2b**). Therefore, we believe that the 2D THE in this work is induced by magnetic skyrmions rather than a superposition of multiple AHE.

**Note S4. Additional analysis of THE and resulted THE-derived skyrmion sizes**

For ferromagnetic materials with skyrmion lattice, the Hall resistivity ($\rho_{xy}$) can be decomposed into $\rho_{OH}$, $\rho_{AH}$ and $\rho_{TH}$, which is further expressed by the following formula[19]:

$$\rho_{xy}=R_0\mu_0H+R_sM+\rho_{TH} \tag{2}$$

where $R_0$ is the ordinary Hall coefficient, $R_s$ the anomalous Hall coefficient, M the magnetization, and $\rho_{TH}$ the topological Hall resistivity.

Further, the $\rho_{TH}$ induced by static skyrmions can be evaluated by the following

formula[19]:

$$\rho_{TH}=PR_0B_{eff}=PR_0n_{sk}\varnothing_0 \qquad (3)$$

where P is the spin polarization of carriers, $R_0$ the ordinary Hall coefficient, $B_{eff}$ an effective magnetic field generated by the skyrmions, $n_{sk}$ the 2D skyrmion density (assuming uniform, regular 2D skyrmion lattices, and each skyrmion carries a topological charge |Q|=1), and $\varnothing_0$ magnetic flux quantum ($\varnothing_0$=h/e, where h is the Plank constant and e is the electronic charge). To extract the $R_0$, the effect of $\rho_{xx}$ on the $\rho_{xy}$ is eliminated by the antisymmetric procedure:

$$\rho'_+(H)=[\rho_+(H)-\rho_-(-H)]/2 \qquad (4)$$

$$\rho'_-(H)=[\rho_-(H)-\rho_+(-H)]/2 \qquad (5)$$

Since the value of P have been recently calculated to be 67% at 10 K and 55% at 300 K in 2D $Fe_3GaTe_2$-based magnetic tunneling junctions[20], the single skyrmion size ($n_{sk}^{-1/2}$) of each 2D O-FGaT/FGaT heterostructure can be roughly derived (**Table S1**) if we reasonably suppose 2D $Fe_3GaTe_{2-x}$ has similar spin polarization with $Fe_3GaTe_2$ since they have similar ferromagnetic properties as shown above.

## Note S5. Discussion of Joule heating effect in current-controlled THE at room temperature

Before we study the current-dependent THE, the Joule heating effect should be discussed and excluded, as the increase of temperature may also influence the THE. In this work, we perform the current-controlled THE in 13 and 19 nm O-FGaT/FGaT heterostructures and evaluate the Joule heating effect by recording the longitudinal resistivity ($\rho_{xx}$) and saturated anomalous Hall resistivity ($\rho_{AH}$) at each current density. This evaluation method has also been applied to current tunable THE in other 2D skyrmion systems[21, 22]. As we all know, the Joule heating effect will gradually increase with the increase of current density. Therefore, if the Joule heating effect is dominant, as the current density increases, the $\rho_{xx}$ will increase while the saturated $\rho_{AH}$ will decrease, similar to the metallic nature and temperature-dependent AHE, respectively. However, as shown in **Figure S17**, the fluctuation of $\rho_{xx}$ and saturated $\rho_{AH}$ with the increase of current densities are negligible, implying the negligible effect of Joule heating effect on the current-controlled THE tests. Furthermore, the relationship of j vs $\rho_{TH}$ and j vs $v_d$ in **main text Figure 4c,d** are consistent with the general magneto-transport law of skyrmion motion according to the previous report[22]. Thus, we believe the reduction of $\rho_{TH}$ in this work may attribute to the current-driven skyrmion motion.

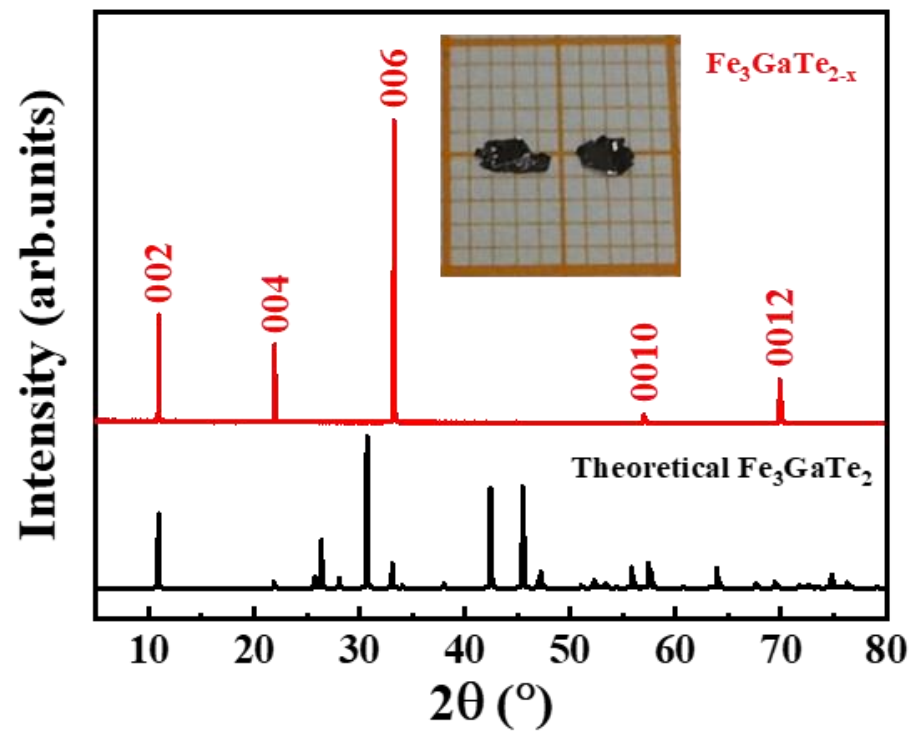

**Figure S1. Crystal photograph and XRD pattern of the bulk $Fe_3GaTe_{2-x}$ crystals.**

The size of each square in inset are 1×1 mm.

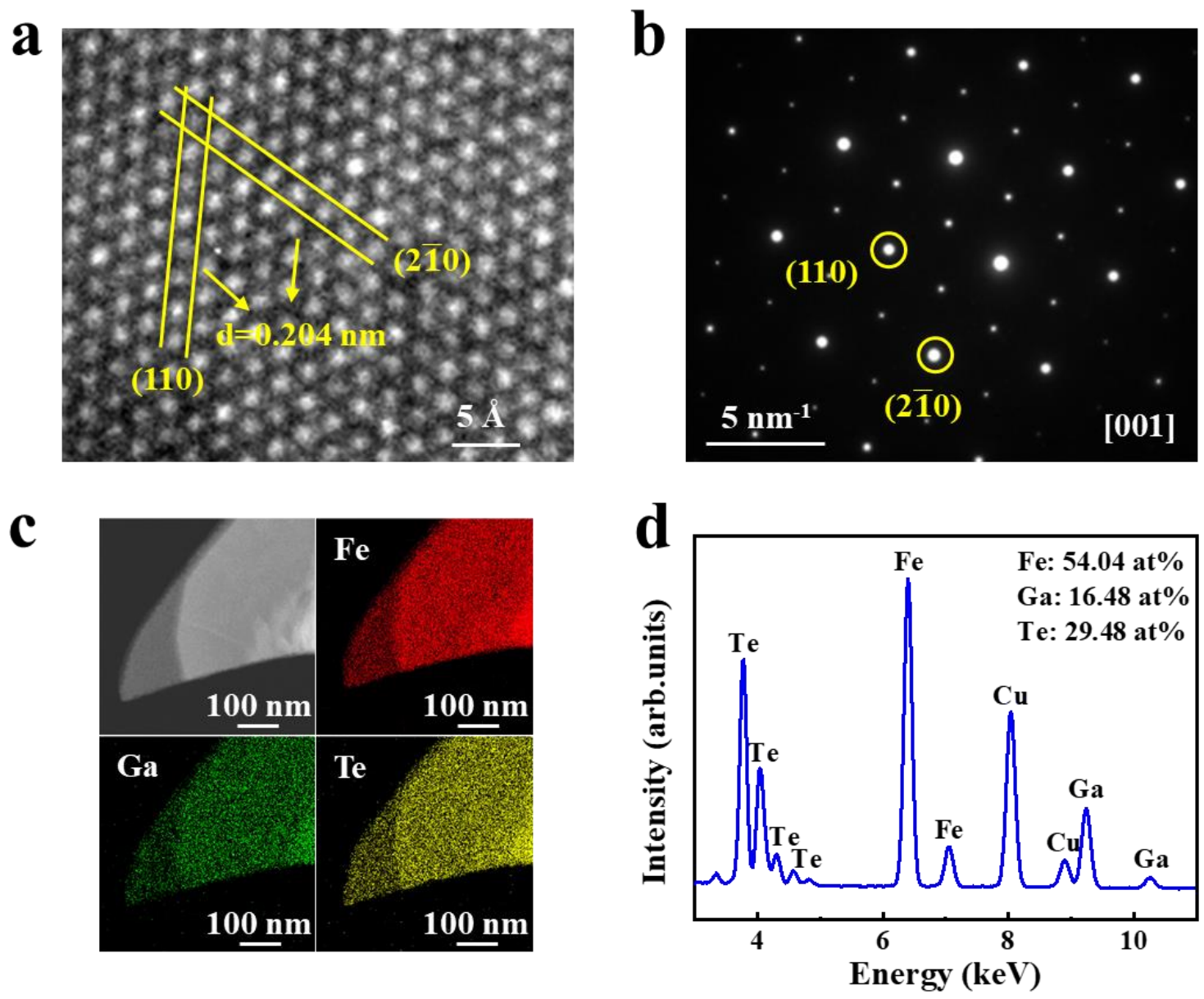

**Figure S2. TEM characterization of a 2D $Fe_3GaTe_{2-x}$ nanosheet along the [001] direction.** (**a,b**) HRTEM image and corresponded SAED pattern of a $Fe_3GaTe_{2-x}$ nanosheet. (**c,d**) EDS elemental mapping images and corresponded EDS spectrum of Fe, Ga, Te.

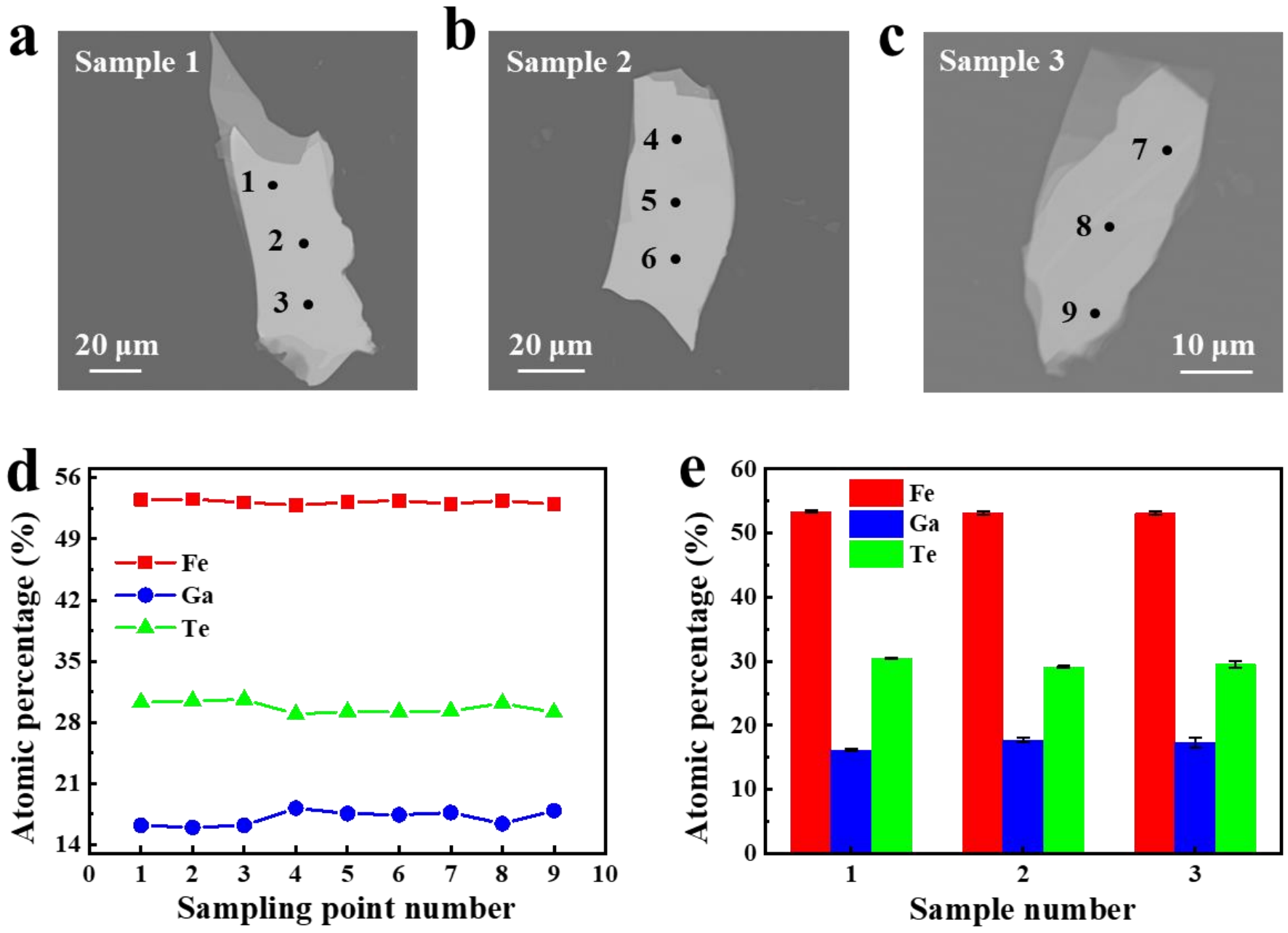

**Figure S3. Quantitative analysis of Fe, Ga and Te content in three pristine $Fe_3GaTe_{2-x}$ single-sheet nanosheets by EPMA.** (**a-c**) EPMA images of three as-tested $Fe_3GaTe_{2-x}$ nanosheets on $SiO_2$/Si substrate. The black points are sampling point. (**d**) Atomic percentages of Fe, Ga, and Te for each sampling point. (**e**) Average atomic percentages of Fe, Ga, and Te for sample 1 (Fe:Ga:Te=3.09:0.93:1.76), sample 2 (Fe:Ga:Te=3.11:1.04:1.70), and sample 3 (Fe:Ga:Te=3.11:1.01:1.73). Error bars s.d., N=3. These results demonstrate the existence of ~15 at% Te vacancies in the $Fe_3GaTe_{2-x}$ crystal. Notably, before the formal EPMA test, Fe-contained, Ga-contained and Te-contained standard samples are used for calibration. For each $Fe_3GaTe_{2-x}$ nanosheet, three positions are randomly selected and each is tested once.

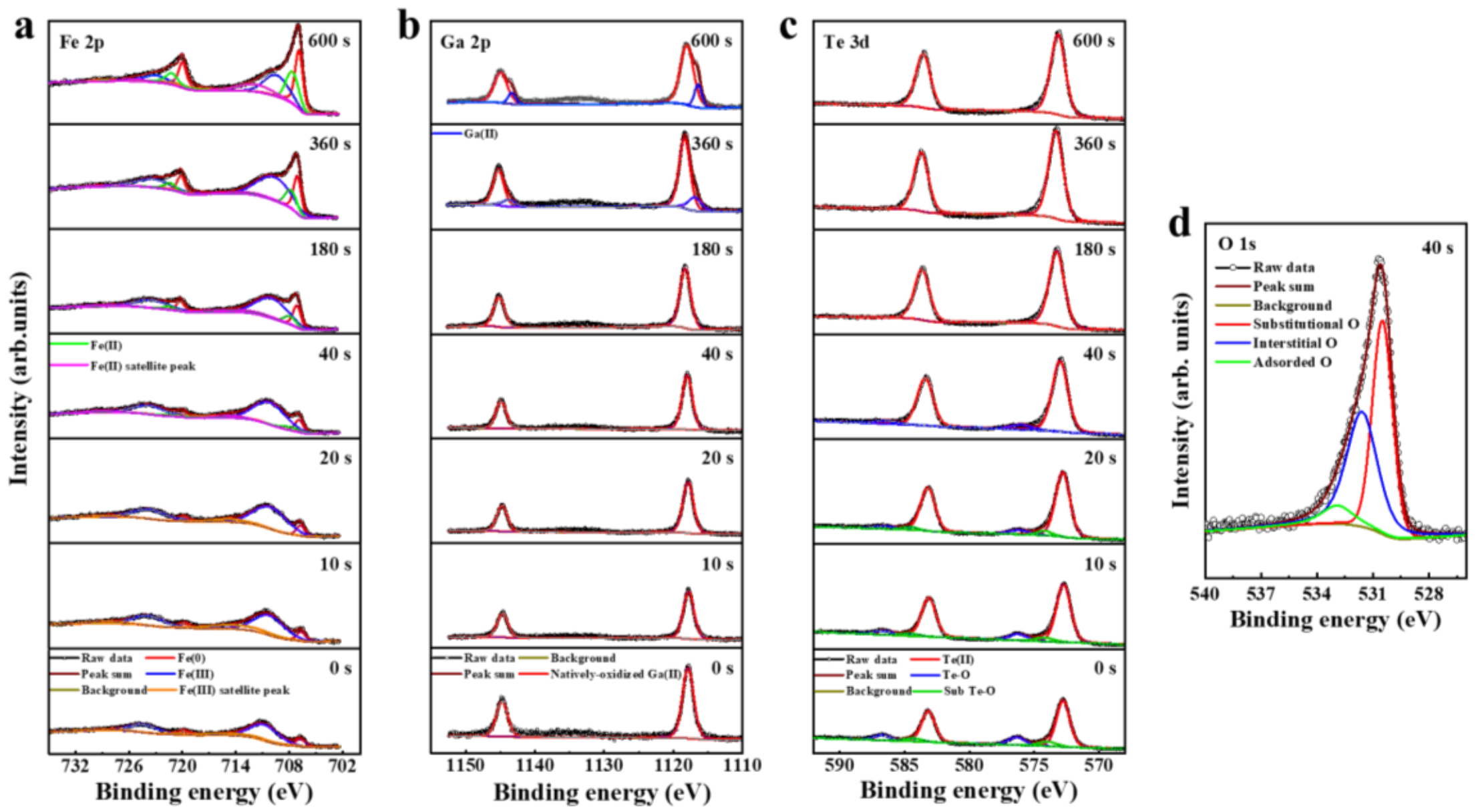

**Figure S4. XPS analysis on an O-FGaT/FGaT surface under different etching time.** (**a**) Fe 2p, (**b**) Ga 2p and (**c**) Te 3d. In order to present the relationship between etching time and intensity, a series of graphs for each element keep a same range of intensity scale. (**d**) Decomposed O 1s spectra. Note that the C 1s (285 eV) used to calibrate the peak position.

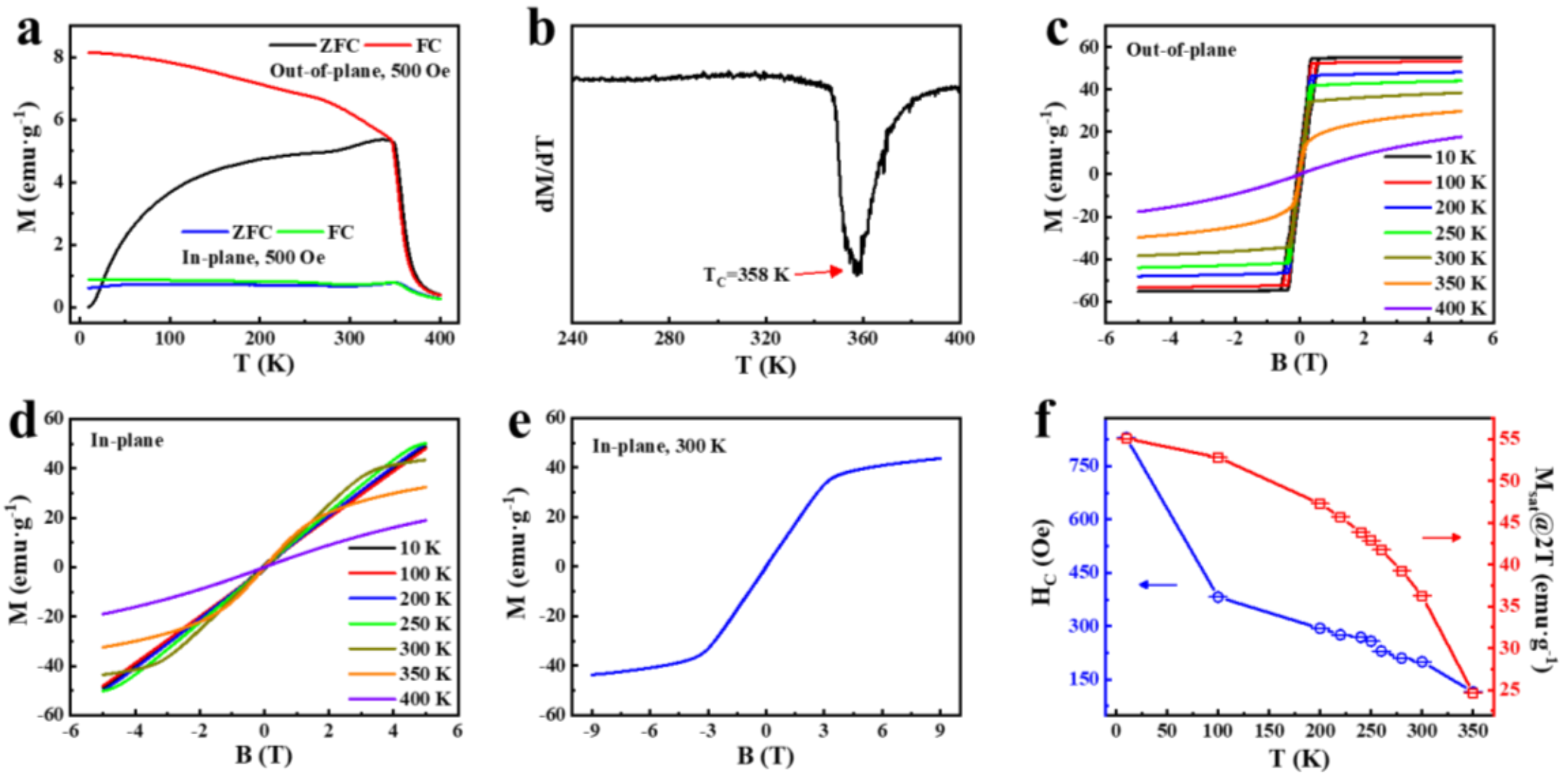

**Figure S5. Above-room-temperature strong ferromagnetism in pristine vdW bulk $Fe_3GaTe_{2-x}$ crystals.** (**a**) Temperature-dependent ZFC-FC curves (M-T) under out-of-plane and in-plane magnetic field. (**b**) First derivative of the out-of-plane ZFC curve. The red arrow shows the ferromagnetic-paramagnetic transition. (**c,d**) M-B curves under varying temperatures with out-of-plane and in-plane magnetic field. (**e**) Room-temperature M-B curve for bulk $Fe_3GaTe_{2-x}$ crystals with in-plane magnetic field from -9 to 9 T. The saturated field $B_{sat}$ is ~3.5 T. (**f**) Temperature-dependent $H_C$ and $M_{sat}$ extracted from (**c**). Error bars sd., N=200.

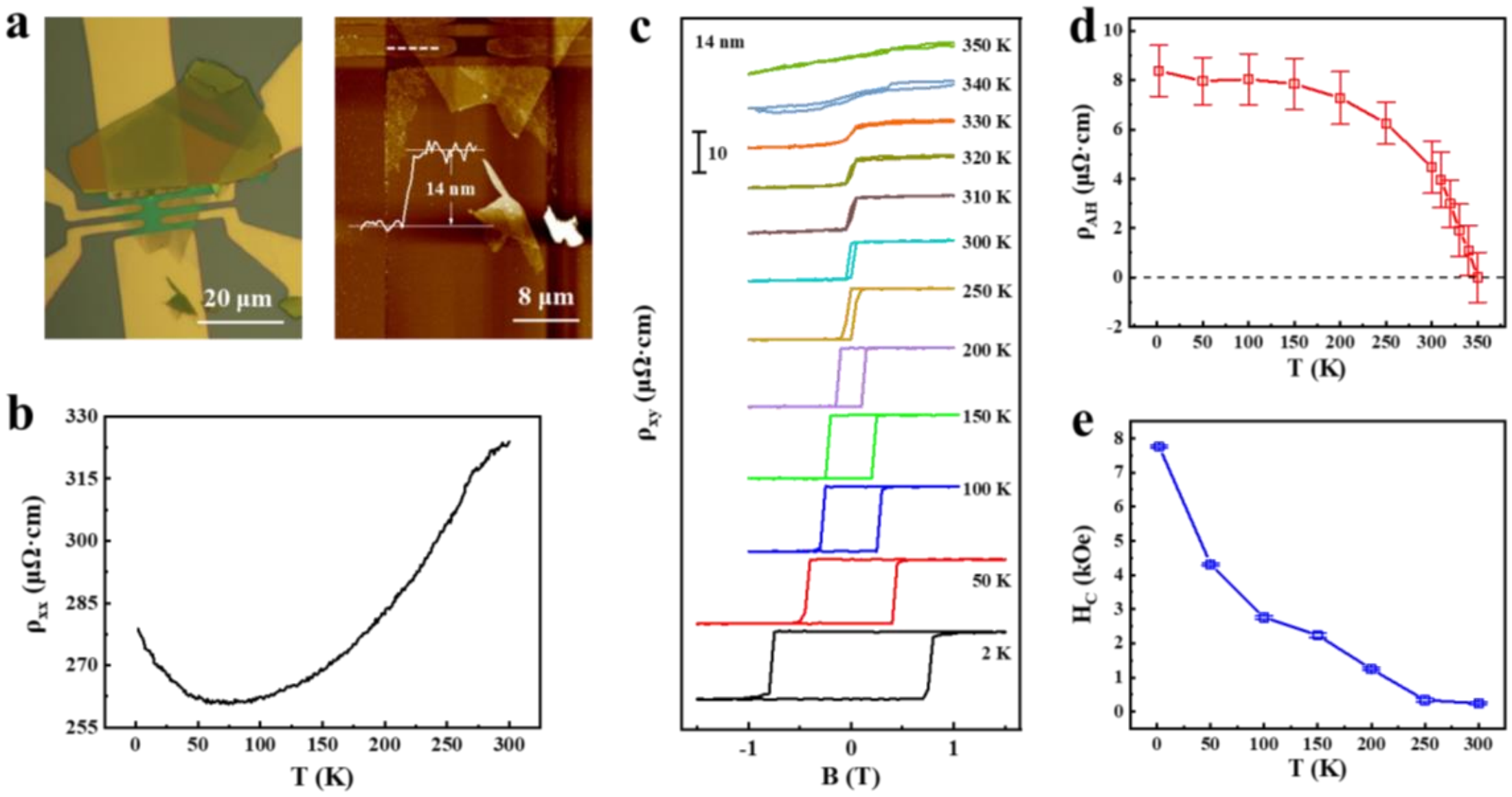

**Figure S6. Magneto-transport measurement of a non-oxidized pristine 14 nm $Fe_3GaTe_{2-x}$ nanosheet.** (**a**) Optical and AFM images of a Hall device based on a $Fe_3GaTe_{2-x}$ nanosheet. (**b**) Temperature-dependent longitudinal resistivity ($\rho_{xx}$) curve. The room temperature $\rho_{xx}$ of this 14 nm nanosheet is $3.24\times10^2$ µΩ·cm. (**c**) AHE under different temperatures. The $T_C$ is determined as ~350 K. (**d**,**e**) Temperature-dependent $\rho_{AH}$ and $H_C$ extracted from (**c**). Error bars sd., N=25 for $\rho_{AH}$ and N=3 for $H_C$.

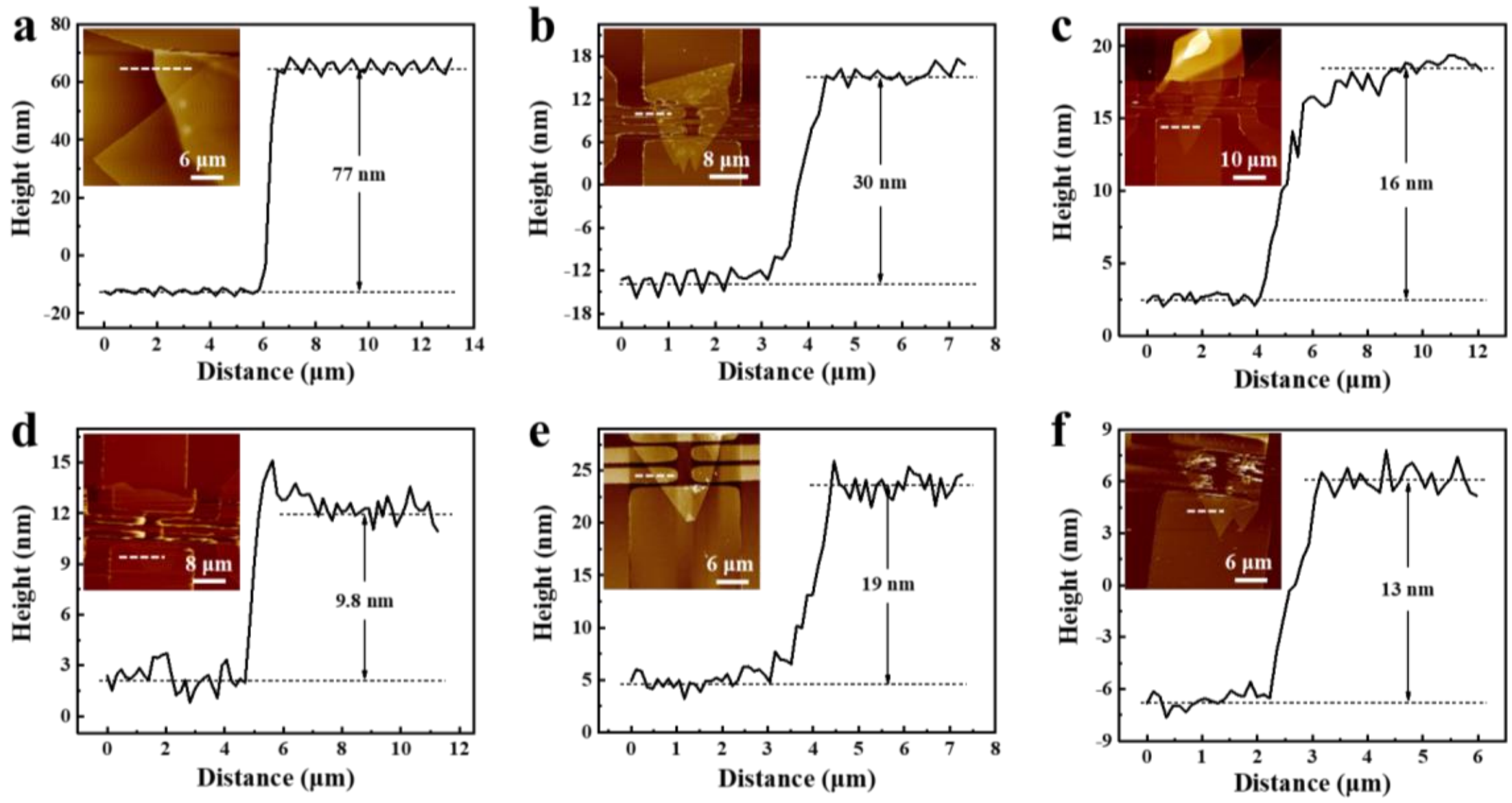

**Figure S7. AFM images and the corresponded profile height of six as-tested Hall devices based on 2D O-FGaT/FGaT heterostructures.**

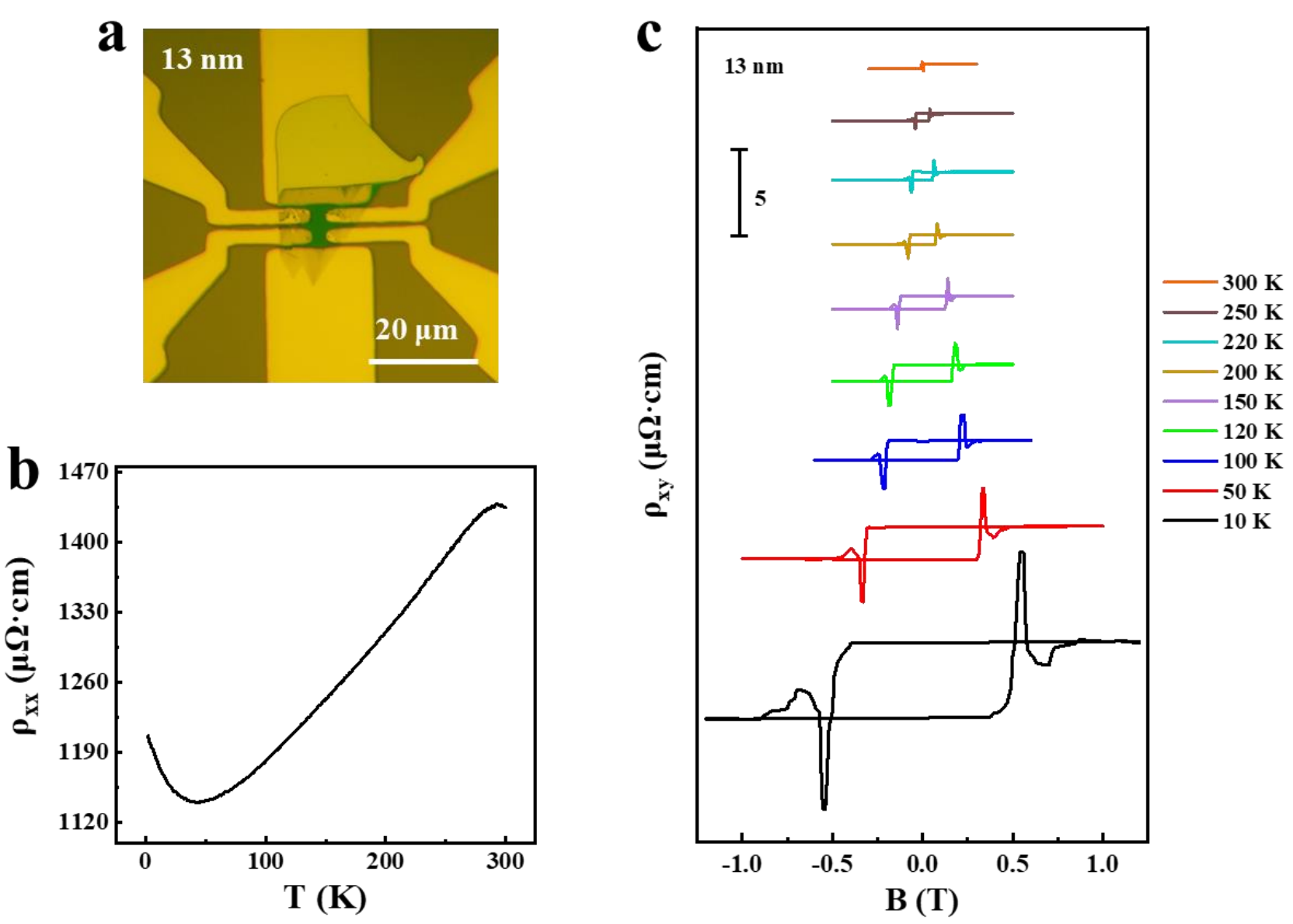

**Figure S8. Magneto-transport measurement of a 13 nm 2D O-FGaT/FGaT heterostructure.** (**a**) Optical image of the Hall device. (**b**) $\rho_{xx}$-T curve. Note that the 2D O-FGaT/FGaT still exhibits metallic nature. This is because oxidation occurs mainly on the upper surface of the sample exposed to air rather than the lower surface in contact with the electrodes. (**c**) $\rho_{xy}$-B curves at varying temperatures.

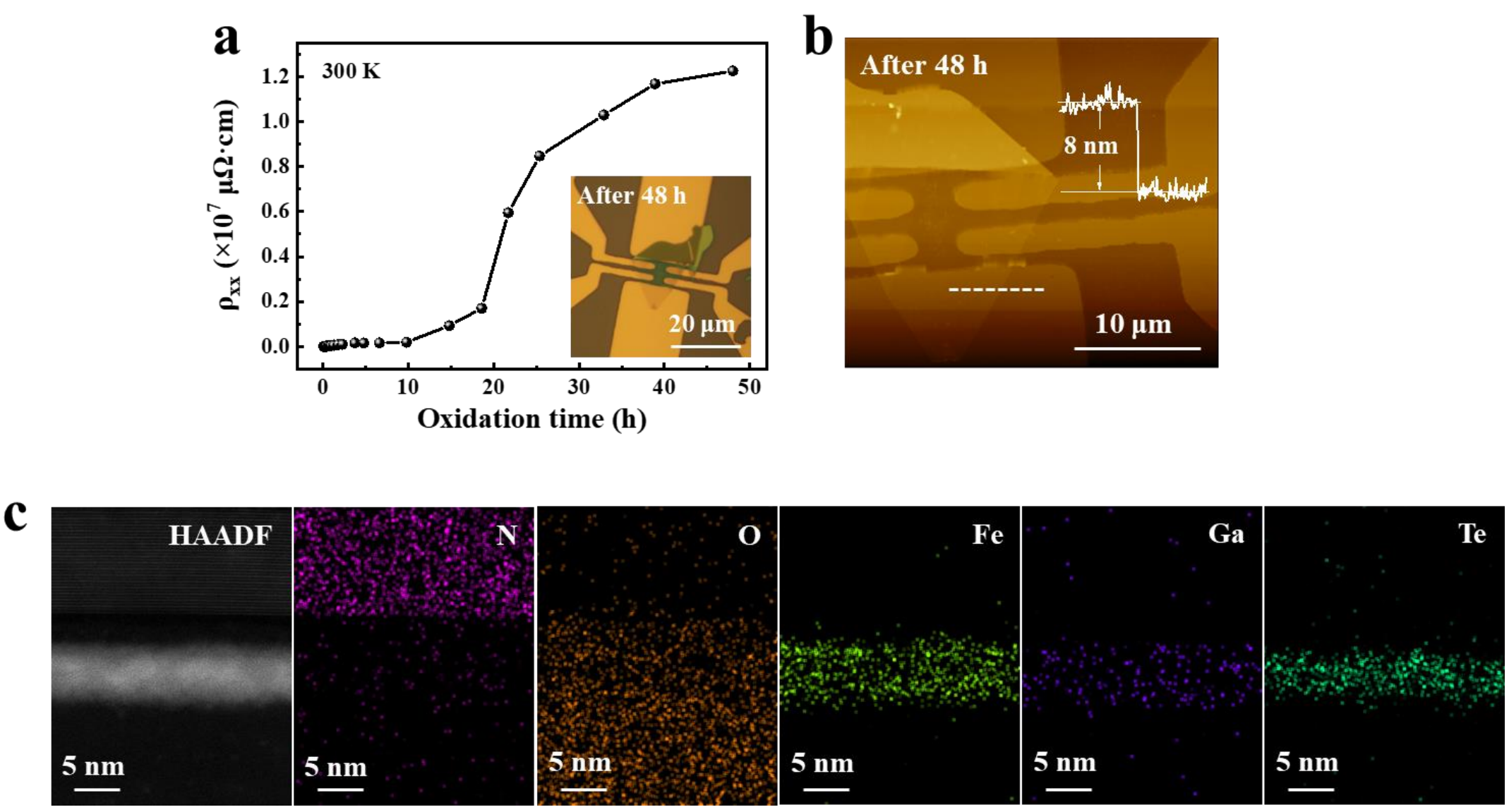

**Figure S9. Natural total oxidation of the 2D $Fe_3GaTe_{2-x}$ nanosheets in the air for 48 h.** (**a**) $\rho_{xx}$ as a function of oxidation time in the air. Inset shows the optical image of a $Fe_3GaTe_{2-x}$ nanosheet after 48 h air oxidation. (**b**) Corresponded AFM image and profile height along the dash line. After the 48 h air oxidation, the thickness of this $Fe_3GaTe_{2-x}$ nanosheet is 8 nm. (**c**) Cross-Sectional HAADF image and corresponded EDS elemental mapping of the 2D $Fe_3GaTe_{2-x}$ nanosheet. The 48 h air oxidation is enough for totally oxidizing the $Fe_3GaTe_{2-x}$ nanosheet, forming a ~8 nm O-FGaT layer.

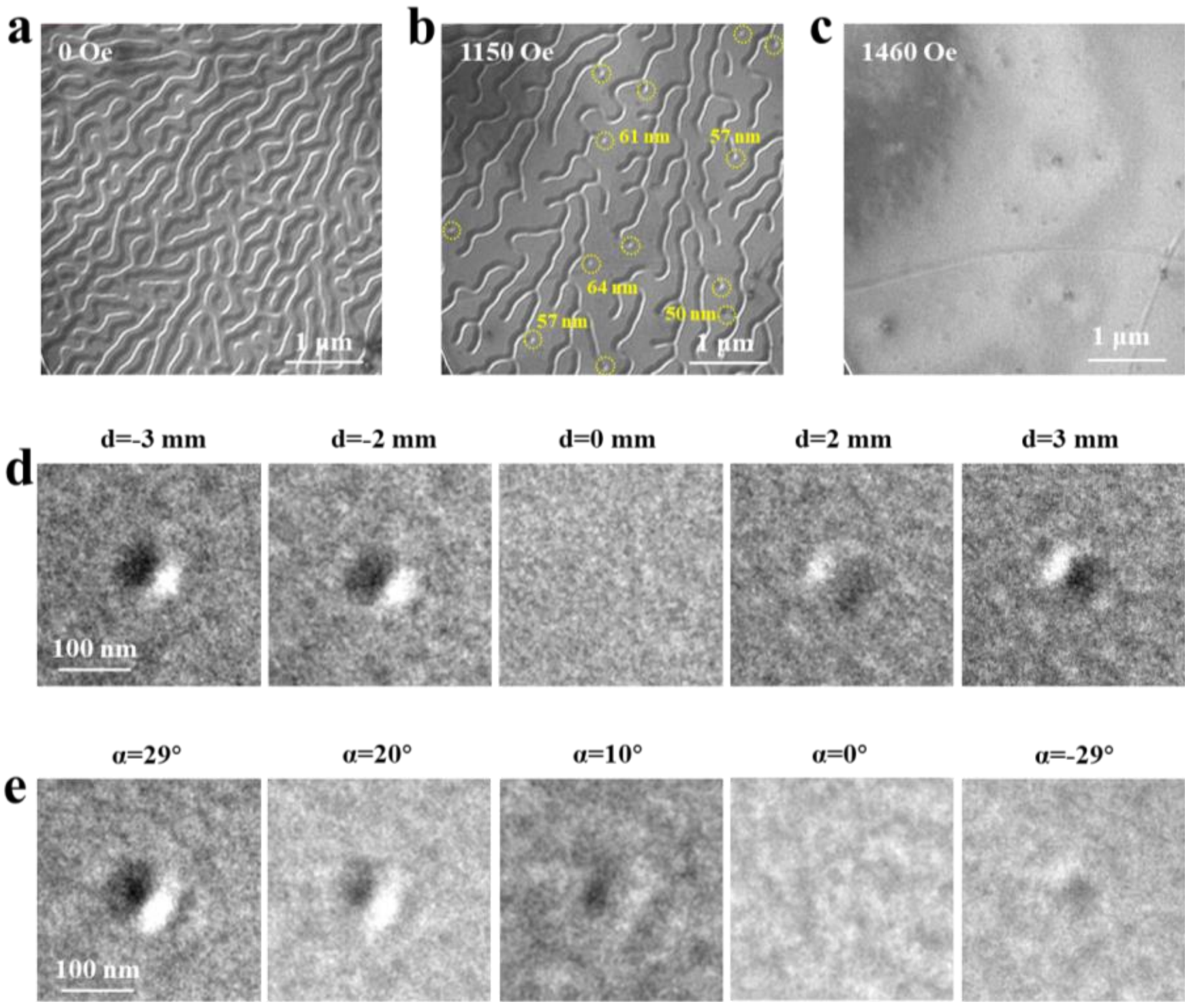

**Figure S10. Observation of Néel-type skyrmions in 2D O-FGaT/FGaT by Lorentz-TEM with the perpendicular magnetic field at 300 K.** (**a-c**) Magnetic-field-driven evolution from stripe domains to skyrmions at α=29°, d=-3 mm, where α is the angle between the sample plane and xy plane, d is the focus distance which positive represents the over-focus and negative represents the under-focus. (**d**) Single skyrmion under different d from under-focus to over-focus at α=29°, B=1400 Oe. (**e**) Single skyrmion under different α from 29° to -29 °at B=1360 Oe, d=-3 mm.

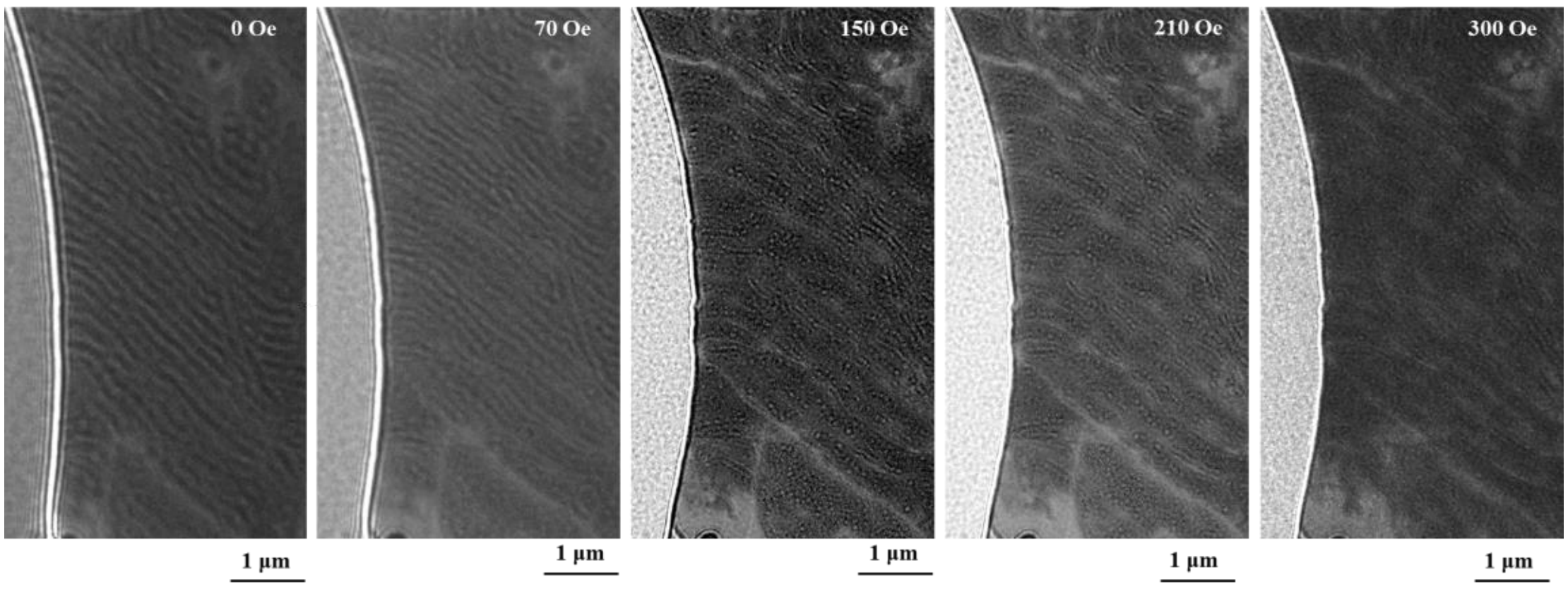

**Figure S11. Magnetic-field-driven evolution from stripe domains to skyrmions in a thin 2D O-FGaT/FGaT by Lorentz-TEM with perpendicular magnetic field at 300 K.** The images are taken at α=19° and d=-2 mm. The magnetic fields perpendicular to the sample are calculated as 0, 66, 142, 198 and 284 Oe.

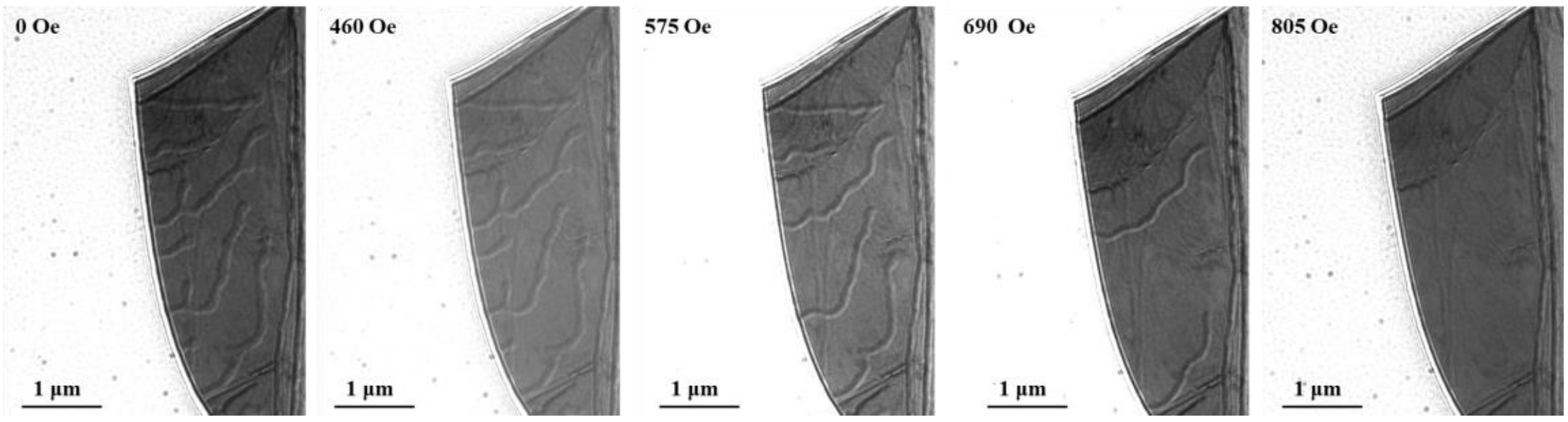

**Figure S12. Lorentz-TEM images under different perpendicular magnetic field in a pristine non-oxidized thin 2D $Fe_3GaTe_{2-x}$ nanosheet at 300 K under α=29°, d=-2 mm.**

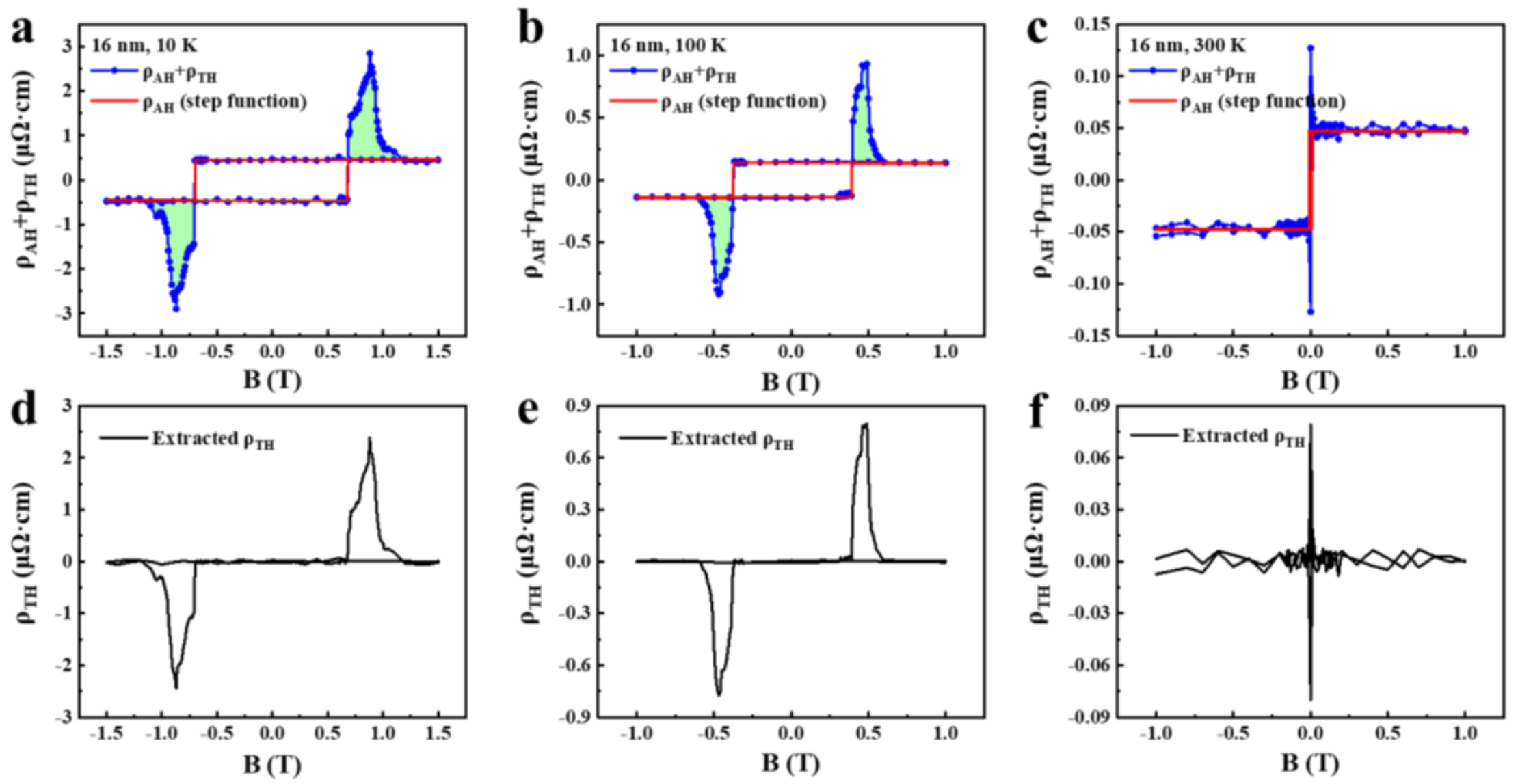

**Figure S13. Extracting the THE signals in 2D O-FGaT/FGaT by step function.** (**a-c**) ($\rho_{AH}$+$\rho_{TH}$) vs B curves at different temperatures. Contributions from AHE and THE terms are marked by red solid lines and light green area, respectively. (**d-f**) Corresponded $\rho_{TH}$ vs B curves extracted from **a-c**.

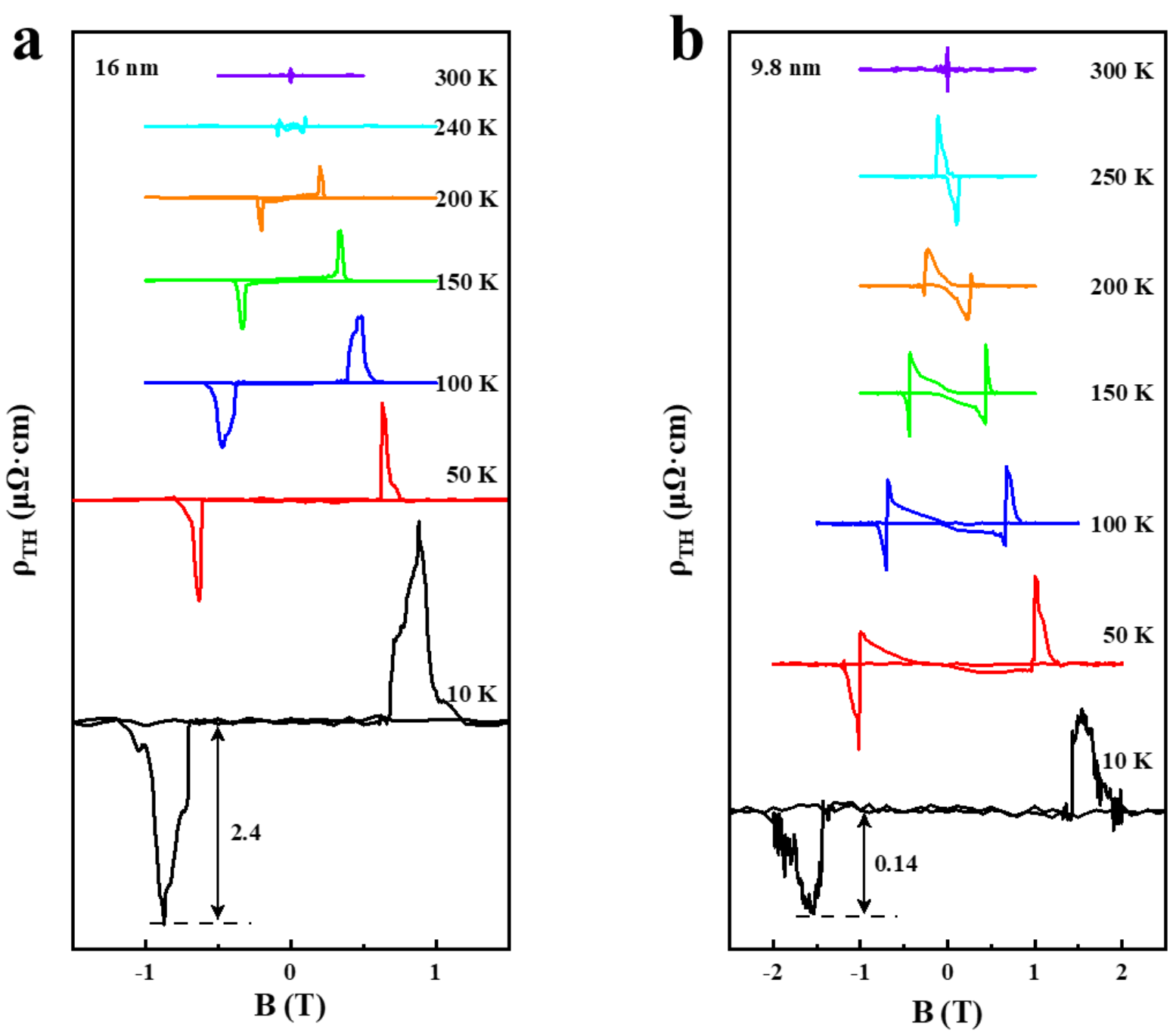

**Figure S14. Temperature-dependent $\rho_{TH}$-B curves for two 2D O-FGaT/FGaT heterostructures with different thickness.** (**a**) 16 nm. (**b**) 9.8 nm.

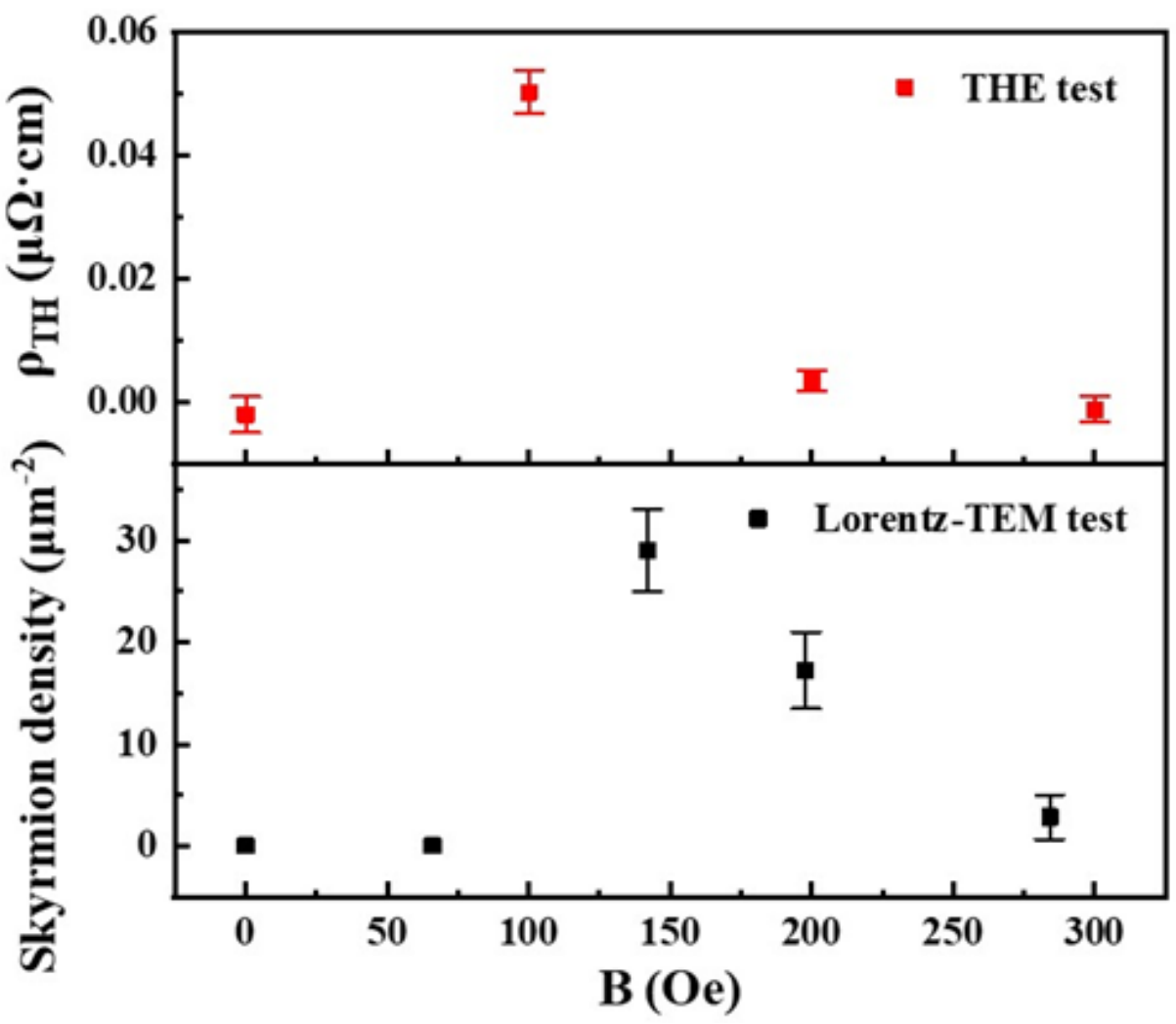

**Figure S15. Magnetic field dependence of the skyrmion density and $\rho_{TH}$ at room temperature in 2D O-FGaT/FGaT.** The skyrmion density in each magnetic field is extracted from three randomly-selected 2 μm×2 μm regions in **Figure S11**. Error bars sd., N=3. The $\rho_{TH}$ in each magnetic field is extracted from the 19 nm O-FGaT/FGaT in **main text Figure 3e**. Error bars sd., N=25. Note that the samples used here for THE test and Lorentz-TEM test are not the same sample.

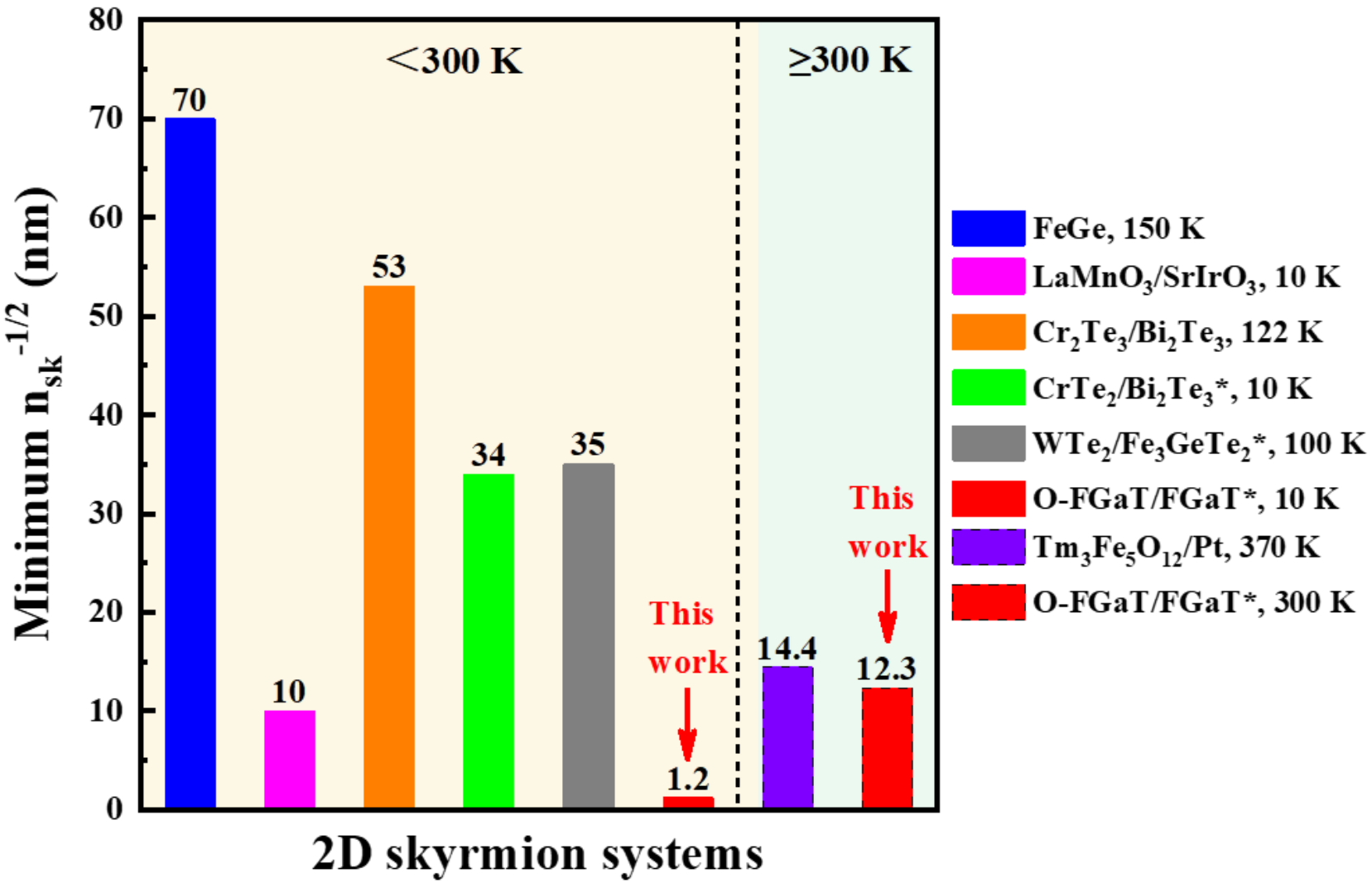

**Figure S16. Comparison of THE tests-derived minimum skyrmion size ($n_{sk}^{-1/2}$) in various 2D skyrmion systems**[19, 23-27]**.** The [*] is the 2D vdW ferromagnet-based skyrmion systems.

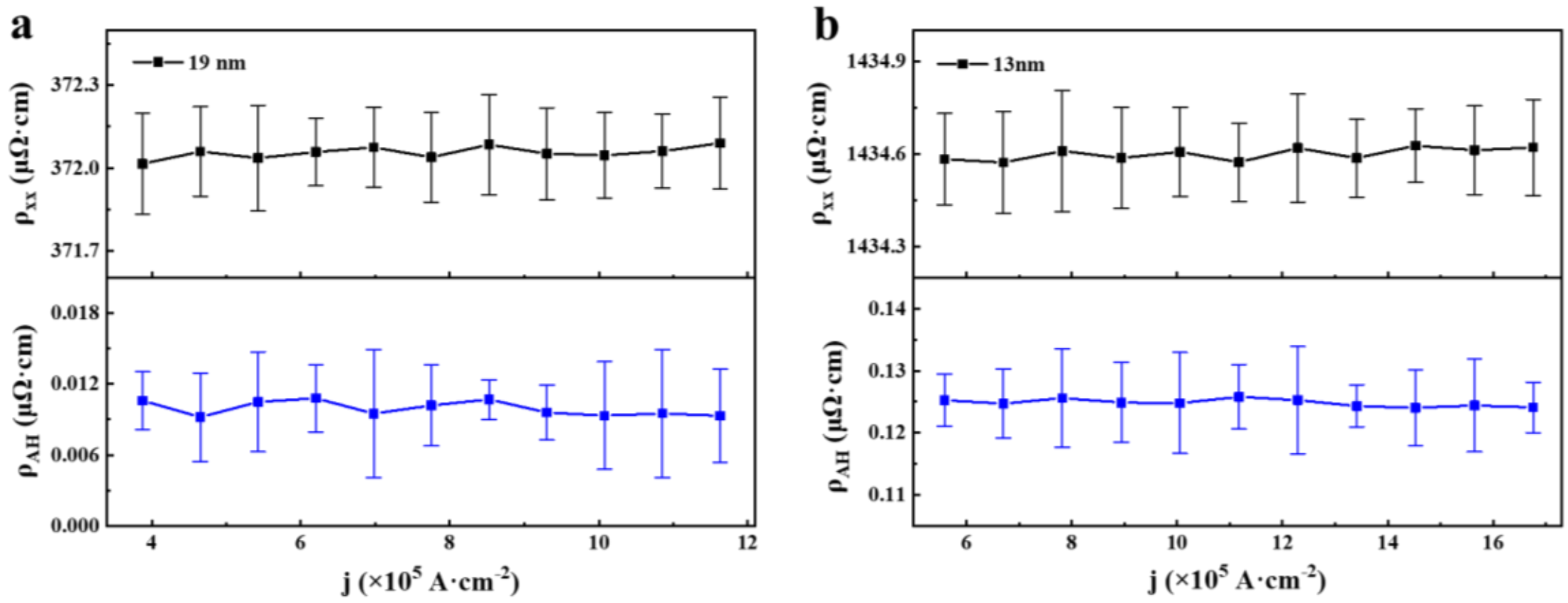

**Figure S17. $\rho_{xx}$ and saturated $\rho_{AH}$ as a function of current densities in 2D O-FGaT/FGaT at room temperature.** (**a**) 19 nm. (**b**) 13 nm. Error bars sd., N=25.

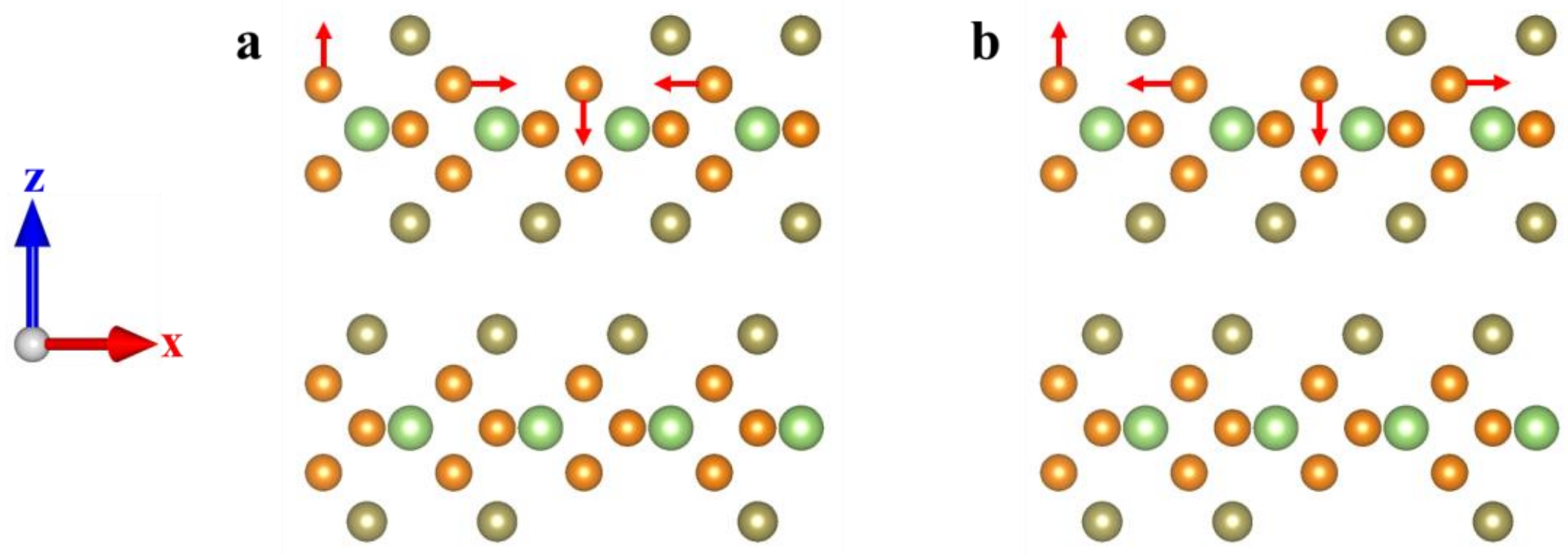

**Figure S18. Theoretical model for calculating DMI in $Fe_3GaTe_{2-x}$.** (**a,b**) Clockwise (CW) (**a**) and anticlockwise (ACW) (**b**) spin configurations of a bilayer $Fe_3GaTe_{2-x}$ for calculating the layer-resolved $d^L$ DMI parameters. The CW (ACW) spin configurations of a single layer Fe atoms are schematically shown by arrows. The spin of other Fe atoms points along the y-axis direction. The orange, green, brown balls denote Fe, Ga, Te atoms, respectively.

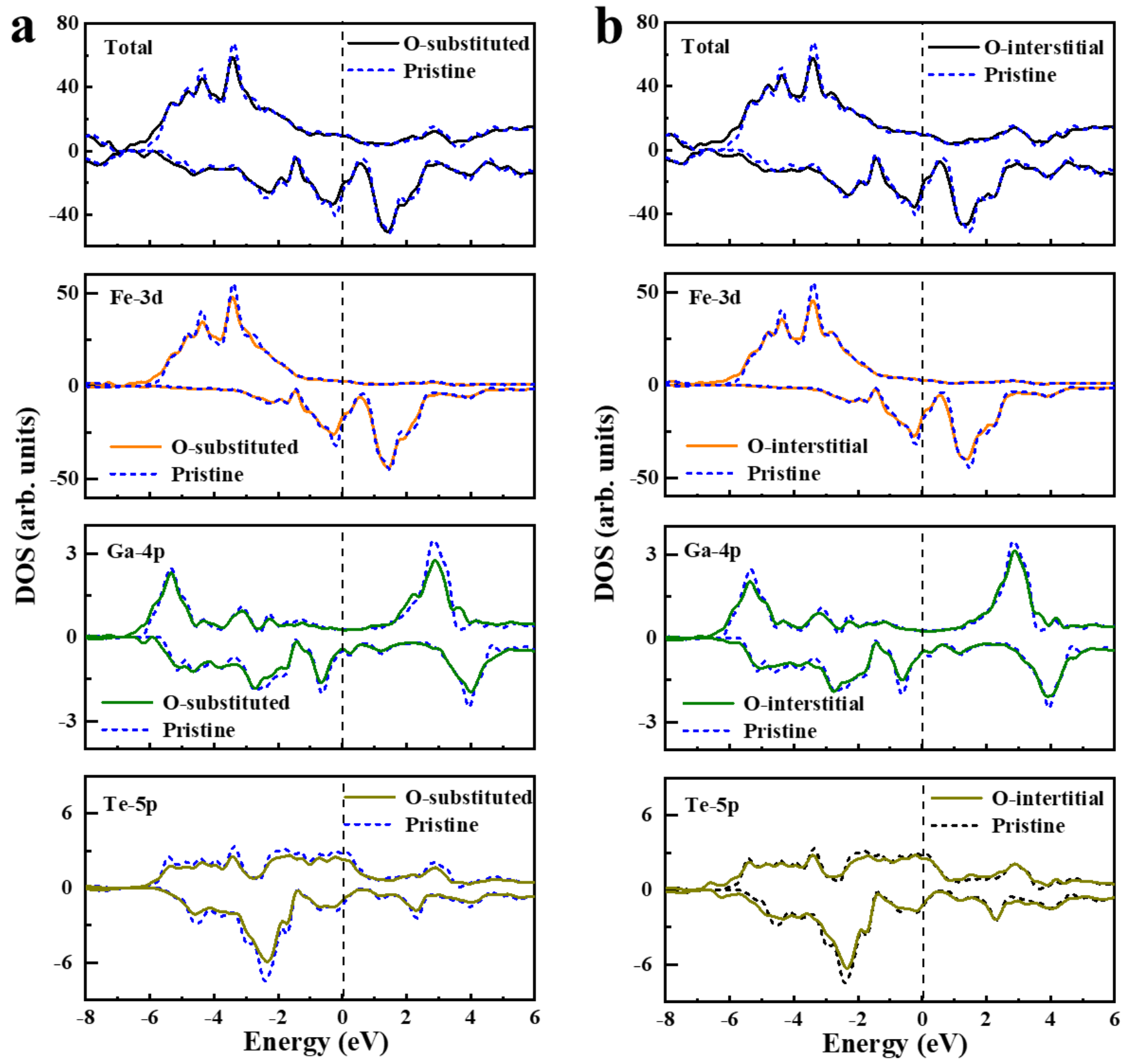

**Figure S19. The density of states (DOSs) comparison of total, Fe-3d, Ga-4p, and Te-5p in oxidized bilayer $Fe_3GaTe_{2-x}$ with that of pristine bilayer $Fe_3GaTe_{2-x}$.** (**a**) O-substituted. (**b**) O-interstitial. The vertical dash lines denote the position of Fermi level. Note that DOSs of total, Fe-3d, Ga-4p, Te-5p all show the shift toward the low-energy direction after introducing surface oxygen atoms in two cases.

**Table S1. Magneto-transport parameters for THE calculation in 2D O-FGaT/FGaT heterostructures with different thickness.**

| Thickness (nm) | T (K) | $\rho_{TH}$ ($\mu\Omega\cdot cm$) | $R_0$ ($\Omega\cdot m\cdot T^{-1}$) | n ($cm^{-3}$) | $n_{sk}^{-1/2}$ (nm) |
|---|---|---|---|---|---|
| 19 | 2 | 1.08 | $1.05\times10^{-11}$ | $5.94\times10^{23}$ | 1.6 |
| | 300 | 0.05 | $7.95\times10^{-11}$ | $7.85\times10^{22}$ | 19 |
| 16 | 10 | 2.4 | $1.56\times10^{-11}$ | $4\times10^{23}$ | 1.4 |
| | 300 | 0.08 | $8.53\times10^{-11}$ | $7.32\times10^{22}$ | 15.6 |
| 13 | 10 | 5.4 | $2.72\times10^{-11}$ | $2.29\times10^{23}$ | 1.2 |
| | 300 | 0.15 | $9.97\times10^{-11}$ | $6.26\times10^{22}$ | 12.3 |
| 9.8 | 10 | 0.14 | $3.65\times10^{-11}$ | $1.71\times10^{23}$ | 8.5 |
| | 300 | 0.03 | $1.31\times10^{-10}$ | $4.76\times10^{22}$ | 31.5 |

**Note:** $\rho_{TH}$ is the THE resistivity, $R_0$ is the ordinary Hall coefficient, n is the carrier concentration, $n_{sk}^{-1/2}$ is the THE tests-derived skyrmion size.

**Table S2. Comparison of the THE temperature window and maximum $\rho_{TH}$ in various 2D skyrmion systems from the literatures.**

| Materials | Temperature window (K) | Maximum $\rho_{TH}$ (μΩ·cm) | Conditions for maximum $\rho_{TH}$ | Ref. |
|---|---|---|---|---|
| FeGe | 10-250 | 0.16 | 18 nm, 150 K | [23] |
| MnSi | 2-50 | 0.025 | 20 nm, 40 K | [28] |
| $Mn_2RhSn$ | 15-200 | 0.26 | 60 nm, 140 K | [29] |
| $Cr_2Te_3/Bi_2Te_3$ | 2-95 | 0.4 | 5.5/19.8 nm, 122 K | [25] |
| $Cr_2Te_3/Cr_2Se_3$ | 10-75 | 0.25 | 10/20 nm, | [17] |
| $Bi_2Se_3/BaFe_{12}O_{19}$ | 2-80 | 0.23 | 6/5 nm, 2 K | [30] |
| $LaMnO_3/SrIrO_3$ | 3-50 | 75 | 5/5 u.c., 10 K | [24] |
| $Tm_3Fe_5O_{12}/Pt$ | 350-410 | 0.0046 | 4/3.2 nm, 370 K | [26] |
| $[Ir/Fe/Co/Pt]_{20}$ | - | 0.03 | 1/0.2/0.6/1 nm, 300 K | [31] |
| $[Co/Pt]_5$ | - | 0.01 | 0.5/1 nm, 300 K | [32] |
| $Pt/Cr_2O_3$ | - | 0.00065 | 2/3 nm, 300 K | [33] |
| $Cr_2Ge_2Te_6/Fe_3GeTe_2$* | 2-100 | 0.31 | 30-40/4 nm, 2 K | [34] |
| $CrTe_2/Bi_2Te_3$* | 10-100 | 1.39 | 6/- nm, 10 K | [19] |
| $WTe_2/Fe_3GeTe_2$* | 2-100 | 1.3 | 3.2/1.1 nm, 2 K | [27] |
| $Sb_2Te_3/Sb_{1.9}V_{0.1}Te_3$* | 6-40 | 0.48 | 3/5 QLs, 6 K | [35] |
| **O-FGaT/FGaT*** | **2-300** | **5.4** | **13 nm, 10 K** | **This work** |
| | | **0.15** | **13 nm, 300 K** | |

**Note:** [*] is the 2D vdW ferromagnet-based skyrmion systems.

**Table S3. Comparison of critical current density ($j_c$) and maximum drift velocity ($v_d$) in various room-temperature 2D skyrmion systems from literatures.**

| Materials | Thickness (nm) | $j_c$ ($A\cdot cm^{-2}$) | Maximum $v_d$ ($m\cdot s^{-1}$) | Ref. |
|---|---|---|---|---|
| $[Pt/Co/Ta]_{15}$ | 3/0.9/4 | $2\times10^7$ | 50 | [36] |
| $Ta/Co/[Pt/Ir/Co]_{10}/Pt$ | 15/0.8/1/1/0.8/3 | $2.7\times10^7$ | 20 | [37] |
| Ta/Pt/Co/MgO/Ta | 3/3/0.97/0.9/2 | $1.3\times10^6$ | 110 | [38] |
| Pt/CoGd/Ta | 6/5/3 | $7.5\times10^7$ | 500 | [39] |
| Pt/CoGd/W | 6/5/3 | $8\times10^7$ | 610 | [39] |
| $[Pt/GdFeCo/MgO]_{20}$ | 3/5/1 | $1\times10^7$ | 50 | [40] |
| $Ta/CoFeB/TaO_x$ | 5/1.2/5 | $1.39\times10^6$ | 0.45 | [41] |
| $[Pt/CoFeB/MgO]_{15}$ | 4.5/0.7/1.4 | $2\times10^7$ | 100 | [36] |
| **O-FGaT/FGaT*** | **13** | $\mathbf{7.82\times10^5}$ | **0.82** | **This work** |
| | **19** | $\mathbf{6.2\times10^5}$ | **0.54** | |

**Note:** [*] is the 2D vdW ferromagnet-based skyrmion systems. The $v_d$ in this Table is taken from the maximum value in each literature.

**Table S4. The binding energies (*$E_b$*) for incorporating oxygen into $Fe_3GaTe_{2-x}$ and $Fe_3GaTe_2$ crystals.**

| | $O_2$ | $Fe_3GaTe_2$ | $O/Fe_3GaTe_2$ | $Fe_3GaTe_{1.7}$ | $O/Fe_3GaTe_{1.7}$ |
|---|---|---|---|---|---|
| $E_b$ (eV) | -9.857 | -139.486 | -149.151 | -134.285 | -146.025 |

**Note:** for $Fe_3GaTe_2$: $E_1=E_{O/Fe_3GaTe_2}-E_{Fe_3GaTe_2}-\frac{1}{2}E_{O_2}=-4.7365$ eV

for $Fe_3GaTe_{1.7}$: $E_2=E_{O/Fe_3GaTe_{1.7}}-E_{Fe_3GaTe_{1.7}}-\frac{1}{2}E_{O_2}=-6.8115$ eV

**Table S5. Average Bader charges (Q) of single atom in pristine, O-substituted, and O-interstitial bilayer $Fe_3GaTe_{2-x}$.** The positive and negative Q values mean that the charges are transferred out of the atoms and transferred into the atoms, respectively.

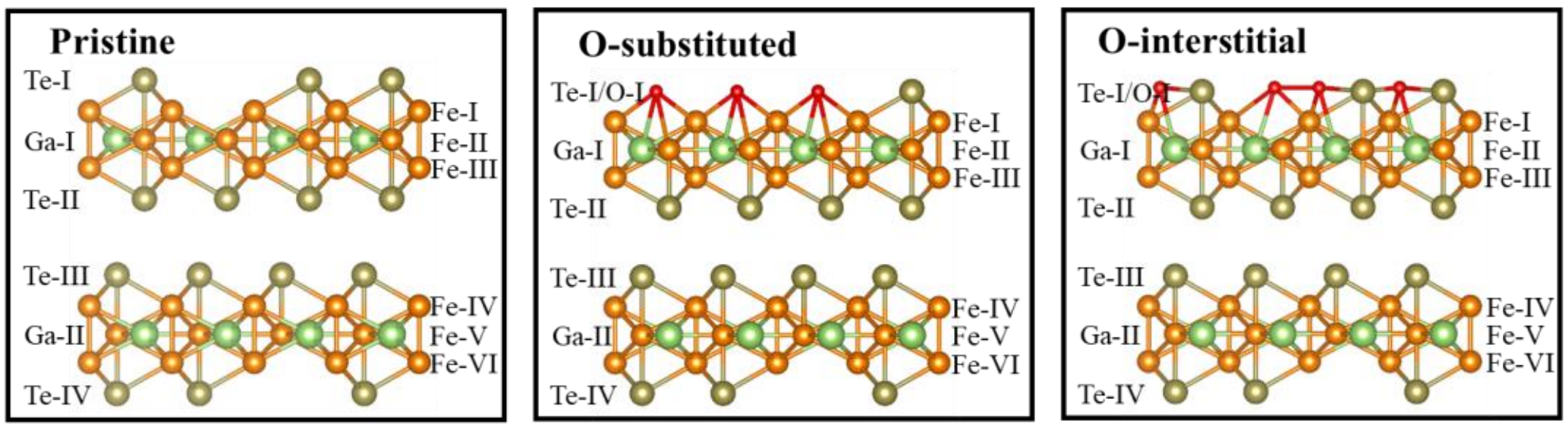

| | | Pristine | O-substituted | O-interstitial |
|---|---|---|---|---|
| Average Q (e) of single atom in each atomic layer of each case | Fe-I | 0.23 | **0.60** | **0.49** |
| | Fe-II | -0.15 | **0.02** | **-0.09** |
| | Fe-III | 0.24 | 0.21 | 0.22 |
| | Fe-IV | 0.23 | 0.25 | 0.25 |
| | Fe-V | -0.15 | -0.14 | -0.17 |
| | Fe-VI | 0.24 | 0.22 | 0.22 |
| | Ga-I | 0.15 | **0.28** | **0.23** |
| | Ga-II | 0.15 | 0.13 | 0.16 |
| | Te-I | -0.28 | **-0.36** | **0.58** |
| | Te-II | -0.25 | -0.25 | -0.26 |
| | Te-III | -0.25 | -0.25 | -0.25 |
| | Te-IV | -0.28 | -0.28 | -0.28 |
| | O-I | - | **-1.03** | **-1.02** |
| Average Q (e) of single atom in each case | Fe | 0.11 | **0.19** | **0.15** |
| | Ga | 0.15 | **0.21** | **0.19** |
| | Te | -0.26 | -0.27 | **-0.08** |
| | O | - | **-1.03** | **-1.02** |

**Note:** The data marked in bold show that the average Q of single atom has significantly changed after introducing O atoms compared with that of pristine case. For the upper panel of this table, the Fe-I, Fe-II, etc. represent the atomic layers shown in upper

images. After the introduction of O atoms, the change of average Q mainly happens in the atomic layers adjacent to O atoms, such as Fe-I, Fe-II, Ga-I and Te-I. For the lower panel of this table, after the introduction of O atoms, the increase of average Q for Fe and Ga atoms and the decrease of average Q for Te atoms all indicate that part of charge is transferred from these atoms to O atoms.

## Supporting References

[1] Kaihang Ye, Kunshan Li, Yirui Lu, Zhongjie Guo, Nan Ni, Hong Liu, Yongchao Huang, Hongbing Ji, P. Wang, *Trends Anal. Chem.* **2019**, 116, 102.

[2] E. Paparazzo, G. M. Ingo, N. Zacchetti, *J. Vac. Sci. Technol. A* **1991**, 9, 1416.

[3] F. Y. Xie, L. Gong, X. Liu, Y. T. Tao, W. H. Zhang, S. H. Chen, H. Meng, J. Chen, *J. Electron. Spectrosc.* **2012**, 185, 112.

[4] Z. Fei, B. Huang, P. Malinowski, W. Wang, T. Song, J. Sanchez, W. Yao, D. Xiao, X. Zhu, A. F. May, W. Wu, D. H. Cobden, J. H. Chu, X. Xu, *Nat. Mater.* **2018**, 17, 778.

[5] C. Gong, L. Li, Z. Li, H. Ji, A. Stern, Y. Xia, T. Cao, W. Bao, C. Wang, Y. Wang, Z. Q. Qiu, R. J. Cava, S. G. Louie, J. Xia, X. Zhang, *Nature* **2017**, 546, 265.

[6] B. Huang, G. Clark, E. Navarro-Moratalla, D. R. Klein, R. Cheng, K. L. Seyler, D. Zhong, E. Schmidgall, M. A. McGuire, D. H. Cobden, W. Yao, D. Xiao, P. Jarillo-Herrero, X. Xu, *Nature* **2017**, 546, 270.

[7] M. Huang, S. Wang, Z. Wang, P. Liu, J. Xiang, C. Feng, X. Wang, Z. Zhang, Z. Wen, H. Xu, G. Yu, Y. Lu, W. Zhao, S. A. Yang, D. Hou, B. Xiang, *ACS Nano* **2021**, 15, 9759.

[8] S. Ikeda, K. Miura, H. Yamamoto, K. Mizunuma, H. D. Gan, M. Endo, S. Kanai, J. Hayakawa, F. Matsukura, H. Ohno, *Nat. Mater.* **2010**, 9, 721.

[9] G. Zhang, F. Guo, H. Wu, X. Wen, L. Yang, W. Jin, W. Zhang, H. Chang, *Nat. Commun.* **2022**, 13, 5067.

[10] X. Zhang, Q. Lu, W. Liu, W. Niu, J. Sun, J. Cook, M. Vaninger, P. F. Miceli, D. J. Singh, S. W. Lian, T. R. Chang, X. He, J. Du, L. He, R. Zhang, G. Bian, Y. Xu, *Nat. Commun.* **2021**, 12, 2492.

[11] G. Kimbell, P. M. Sass, B. Woltjes, E. K. Ko, T. W. Noh, W. Wu, J. W. A. Robinson, *Phys. Rev. Mater.* **2020**, 4, 054414.

[12] K. M. Fijalkowski, M. Hartl, M. Winnerlein, P. Mandal, S. Schreyeck, K. Brunner, C. Gould, L. W. Molenkamp, *Phys. Rev. X* **2020**, 10, 011012.

[13] A. Gerber, *Phys. Rev. B* **2018**, 98, 214440.

[14] D. Kan, T. Moriyama, K. Kobayashi, Y. Shimakawa, *Phys. Rev. B* **2018**, 98, 180408(R).

[15] L. Wang, Q. Feng, H. G. Lee, E. K. Ko, Q. Lu, T. W. Noh, *Nano Lett.* **2020**, 20, 2468.

[16] G. Kimbell, C. Kim, W. Wu, M. Cuoco, J. W. A. Robinson, *Commun Mater* **2022**, 3, 19.

[17] J. H. Jeon, H. R. Na, H. Kim, S. Lee, S. Song, J. Kim, S. Park, J. Kim, H. Noh, G. Kim, S. K. Jerng, S. H. Chun, *ACS Nano* **2022**, 16, 8974.

[18] L. Tai, B. Dai, J. Li, H. Huang, S. K. Chong, K. L. Wong, H. Zhang, P. Zhang, P. Deng, C. Eckberg, G. Qiu, H. He, D. Wu, S. Xu, A. Davydov, R. Wu, K. L. Wang, *ACS Nano* **2022**, 16, 17336.

[19] X. Zhang, S. C. Ambhire, Q. Lu, W. Niu, J. Cook, J. S. Jiang, D. Hong, L. Alahmed, L. He, R. Zhang, Y. Xu, S. S. Zhang, P. Li, G. Bian, *ACS Nano* **2021**, 15, 15710.

[20] W. Zhu, S. Xie, H. Lin, G. Zhang, H. Wu, T. Hu, Z. Wang, X. Zhang, J. Xu, Y. Wang, Y. Zheng, F. Yan, J. Zhang, L. Zhao, A. Patanè, J. Zhang, H. Chang, K. Wang, *Chin. Phys. Lett.* **2022**, 39, 128501.

[21] Hongrui Zhang, David Raftrey, Ying-Ting Chan, Yu-Tsun Shao, Rui Chen, Xiang Chen, Xiaoxi Huang, Jonathan T. Reichanadter, Kaichen Dong, Sandhya Susarla, Lucas Caretta, Zhen Chen, Jie Yao, Peter Fischer, Jeffrey B. Neaton, Weida Wu, David A. Muller, Robert J. Birgeneau, R. Ramesh, *Sci. Adv.* **2022**, 8, eabm7103.
[22] D. Liang, J. P. DeGrave, M. J. Stolt, Y. Tokura, S. Jin, *Nat. Commun.* **2015**, 6, 8217.
[23] S. X. Huang, C. L. Chien, *Phys. Rev. Lett.* **2012**, 108, 267201.
[24] Elizabeth Skoropata, John Nichols, Jong Mok Ok, Rajesh V. Chopdekar, Eun Sang Choi, Ankur Rastogi, Changhee Sohn, Xiang Gao, Sangmoon Yoon, Thomas Farmer, Ryan D. Desautels, Yongseong Choi, Daniel Haskel, John W. Freeland, Satoshi Okamoto, Matthew Brahlek, H. N. Lee, *Sci. Adv.* **2020**, 6, eaaz3902.
[25] J. Chen, L. Zhou, L. Wang, Z. Yan, X. Deng, J. Zhou, J.-w. Mei, Y. Qiu, B. Xi, X. Wang, H. He, G. Wang, *Cryst. Growth. Des.* **2021**, 22, 140.
[26] Q. Shao, Y. Liu, G. Yu, S. K. Kim, X. Che, C. Tang, Q. L. He, Y. Tserkovnyak, J. Shi, K. L. Wang, *Nat. Electron.* **2019**, 2, 182.
[27] Y. Wu, S. Zhang, J. Zhang, W. Wang, Y. L. Zhu, J. Hu, G. Yin, K. Wong, C. Fang, C. Wan, X. Han, Q. Shao, T. Taniguchi, K. Watanabe, J. Zang, Z. Mao, X. Zhang, K. L. Wang, *Nat. Commun.* **2020**, 11, 3860.
[28] T. Yokouchi, N. Kanazawa, A. Tsukazaki, Y. Kozuka, M. Kawasaki, M. Ichikawa, F. Kagawa, Y. Tokura, *Phys. Rev. B* **2014**, 89, 064416.
[29] P. K. Sivakumar, B. Gobel, E. Lesne, A. Markou, J. Gidugu, J. M. Taylor, H. Deniz, J. Jena, C. Felser, I. Mertig, S. S. P. Parkin, *ACS Nano* **2020**, 14, 13463.
[30] P. Li, J. Ding, S. S. Zhang, J. Kally, T. Pillsbury, O. G. Heinonen, G. Rimal, C. Bi, A. DeMann, S. B. Field, W. Wang, J. Tang, J. S. Jiang, A. Hoffmann, N. Samarth, M. Wu, *Nano Lett.* **2021**, 21, 84.
[31] A. Soumyanarayanan, M. Raju, A. L. Gonzalez Oyarce, A. K. C. Tan, M. Y. Im, A. P. Petrovic, P. Ho, K. H. Khoo, M. Tran, C. K. Gan, F. Ernult, C. Panagopoulos, *Nat. Mater.* **2017**, 16, 898.
[32] M. V. Sapozhnikov, N. S. Gusev, S. A. Gusev, D. A. Tatarskiy, Y. V. Petrov, A. G. Temiryazev, A. A. Fraerman, *Phys. Rev. B* **2021**, 103, 054429.
[33] Y. Cheng, S. Yu, M. Zhu, J. Hwang, F. Yang, *Phys. Rev. Lett.* **2019**, 123, 237206.
[34] Y. Wu, B. Francisco, Z. Chen, W. Wang, Y. Zhang, C. Wan, X. Han, H. Chi, Y. Hou, A. Lodesani, G. Yin, K. Liu, Y. T. Cui, K. L. Wang, J. S. Moodera, *Adv. Mater.* **2022**, 34, e2110583.
[35] W. Wang, Y. F. Zhao, F. Wang, M. W. Daniels, C. Z. Chang, J. Zang, D. Xiao, W. Wu, *Nano Lett.* **2021**, 21, 1108.
[36] S. Woo, K. Litzius, B. Kruger, M. Y. Im, L. Caretta, K. Richter, M. Mann, A. Krone, R. M. Reeve, M. Weigand, P. Agrawal, I. Lemesh, M. A. Mawass, P. Fischer, M. Klaui, G. S. Beach, *Nat. Mater.* **2016**, 15, 501.
[37] W. Legrand, D. Maccariello, N. Reyren, K. Garcia, C. Moutafis, C. Moreau-Luchaire, S. Collin, K. Bouzehouane, V. Cros, A. Fert, *Nano Lett.* **2017**, 17, 2703.
[38] R. Juge, K. Bairagi, K. G. Rana, J. Vogel, M. Sall, D. Mailly, V. T. Pham, Q. Zhang, N. Sisodia, M. Foerster, L. Aballe, M. Belmeguenai, Y. Roussigne, S. Auffret, L. D. Buda-Prejbeanu, G. Gaudin, D. Ravelosona, O. Boulle, *Nano Lett.* **2021**, 21, 2989.
[39] Y. Quessab, J. W. Xu, E. Cogulu, S. Finizio, J. Raabe, A. D. Kent, *Nano Lett.* **2022**,

22, 6091.
[40] S. Woo, K. M. Song, X. Zhang, Y. Zhou, M. Ezawa, X. Liu, S. Finizio, J. Raabe, N. J. Lee, S. I. Kim, S. Y. Park, Y. Kim, J. Y. Kim, D. Lee, O. Lee, J. W. Choi, B. C. Min, H. C. Koo, J. Chang, *Nat. Commun.* **2018**, 9, 959.
[41] G. Yu, P. Upadhyaya, Q. Shao, H. Wu, G. Yin, X. Li, C. He, W. Jiang, X. Han, P. K. Amiri, K. L. Wang, *Nano Lett.* **2017**, 17, 261.